\documentclass[useAMS,usenatbib]{mn2e}

\usepackage{amssymb}
\usepackage{times}
\usepackage{graphicx}

\def\bj{b_{\rm J}}

\def\etal{\rm{et al.}}
\def \Halpha {${\rm H} \alpha $}
\def \etapar {$\eta $}

\title[Nature of relative bias in the 2dFGRS]{The 2dF Galaxy Redshift Survey: the nature of the relative bias between galaxies of different spectral type}
\author[Conway et al. (The 2dFGRS Team)]{
\parbox[t]{\textwidth}{Edward Conway$^1$\footnotemark[1],
Steve Maddox$^1$\footnotemark[1], Vivienne Wild$^3$, John A.\ Peacock$^6$, Ed Hawkins$^1$, Peder Norberg$^{5}$, Darren S.\ Madgwick$^{4}$, Ivan K.\ Baldry$^7$, Carlton M.\ Baugh$^2$, Joss Bland-Hawthorn$^8$, Terry
Bridges$^8$, Russell Cannon$^8$, Shaun Cole$^2$, Matthew Colless$^9$, Chris
Collins$^{10}$, Warrick Couch$^{11}$, Gavin Dalton$^{12, 13}$, Roberto
De Propris$^{11}$, Simon P.\ Driver$^9$, George Efstathiou$^3$,
Richard S.\ Ellis$^{14}$, Carlos S.\ Frenk$^2$, Karl Glazebrook$^7$,
Carole Jackson$^9$, Bryn Jones$^1$, Ofer Lahav$^{16}$, Ian Lewis$^{12}$,
Stuart Lumsden$^{15}$, Will Percival$^6$, Bruce A.\ Peterson$^9$, Will Sutherland$^3$ and Keith Taylor$^{14}$ (The 2dFGRS Team)\\ 
\small{\it
$^1$School of Physics \& Astronomy, University of Nottingham,
Nottingham NG7 2RD, UK\\ 
$^2$Department of Physics, University of
Durham, South Road, Durham DH1 3LE, UK \\ 
$^3$Institute of Astronomy,
University of Cambridge, Madingley Road, Cambridge CB3 0HA, UK \\
$^4$Hubble Fellow, Lawrence Berkeley National Laboratory, MS50R-5032, Berkeley, CA 94720, USA\\
$^5$Institut f\"ur Astronomie, ETH H\"onggerberg, CH-8093 Z\"urich,
Switzerland\\
$^6$Institute for Astronomy, University of Edinburgh, Royal Observatory, Blackford Hill, Edinburgh EH9 3HJ, UK\\ 
$^7$Department of Physics \& Astronomy, Johns Hopkins University,
Baltimore, MD 21218-2686, USA \\ 
$^8$Anglo-Australian Observatory,
P.O.\ Box 296, Epping, NSW 2121, Australia\\ 
$^9$Research School of
Astronomy \& Astrophysics, The Australian National University, Weston
Creek, ACT 2611, Australia \\ 
$^{10}$Astrophysics Research Institute, Liverpool John Moores University, Twelve Quays House, Birkenhead, L14 1LD, UK \\ 
$^{11}$Department of Astrophysics, University of New South
Wales, Sydney, NSW 2052, Australia \\ 
$^{12}$Department of Physics,
University of Oxford, Keble Road, Oxford OX1 3RH, UK \\ 
$^{13}$Space Science and Technology Division, Rutherford Appleton Laboratory,
Chilton, Didcot, OX11 0QX, UK \\ 
$^{14}$Department of Astronomy,
California Institute of Technology, Pasadena, CA 91125, USA \\
$^{15}$Department of Physics, University of Leeds, Woodhouse Lane,
Leeds, LS2 9JT, UK\\
$^{16}$Department of Physics and Astronomy, University College London, Gower Street, London WC1E 6BT, UK\\\\}
\footnotemark[1]~E-mail: edward.conway@astro.nottingham.ac.uk (EC), steve.maddox@nottingham.ac.uk (SM) }}
\begin{document}

\date{\today}

\pagerange{\pageref{firstpage}--\pageref{lastpage}} \pubyear{2003}

\maketitle

\label{firstpage}

\begin{abstract}
We present an analysis of the relative bias between early- and
late-type galaxies in the Two-degree Field Galaxy Redshift Survey
(2dFGRS) -- as defined by the \etapar\ parameter of
\citet{Madgwick_2002}, which quantifies the spectral type of galaxies
in the survey. Our analysis examines the joint counts in cells between
early- and late-type galaxies, using approximately cubical cells with
sides ranging from 7$h^{-1}$Mpc to 42$h^{-1}$Mpc. We measure the
variance of the counts in cells using the method of Efstathiou et
al.~(1990), which we find requires a correction for a finite volume
effect equivalent to the integral constraint bias of the
autocorrelation function. Using a maximum likelihood technique we fit
lognormal models to the one-point density distribution, and develop
methods of dealing with biases in the recovered variances resulting
from this technique. We use a modified $\chi^{2}$ technique to
determine to what extent the relative bias is consistent with a simple
linear bias relation; this analysis results in a significant detection
of nonlinearity/stochasticity even on large scales. We directly fit
deterministic models for the joint density distribution function,
$f(\delta_{E},\delta_{L})$, to the joint counts in cells using a
maximum likelihood technique. Our results are consistent with a scale
invariant relative bias factor on all scales studied. Linear bias is
ruled out on scales less than $\ell=28h^{-1}$Mpc. A power-law bias
model is a significantly better fit to the data on all but the largest scales
studied; the relative goodness of fit of this model as compared to
that of the linear bias model suggests that any nonlinearity is
negligible for $\ell\gtrsim40h^{-1}$Mpc, consistent with the
expectation from theory that the bias should become linear on large
scales.
\end{abstract}


\section{Introduction}

Measurements of large-scale structure from galaxy redshift surveys
obviously measure the distribution of luminous matter only; the total
mass distribution will be dominated by dark matter. The question of
how the galaxies trace the total matter density field is therefore
extremely pertinent, both to the estimation of cosmological parameters,
and also as a probe of the physics of galaxy formation. A common
assumption is `linear biasing', which can be expressed
$\delta_{g}=b\delta_{m}$, where $\delta_{g}$, $\delta_{m}$ are the
fractional overdensities relative to the mean in galaxies and mass
respectively. This assumption becomes unphysical when $b >
1$ since, by definition, $\delta_{g} \ge -1$, but we can still define a bias parameter, $b(r)$, by
e.g. $\xi_{gg}(r)=b(r)^{2}\xi_{mm}(r)$. Many of the constraints on cosmological parameters derived from large-scale structure measurements rely on an understanding of galaxy bias. Both the 2dFGRS power spectrum analysis (Percival \etal~2001) and the constraints obtained for the neutrino mass (Elgar\o y \& Lahav 2003) assume scale independent bias. Joint constraints obtained by combining the 2dFGRS results with measurements of the CMB power spectrum (Percival et al.\ 2002; Efstathiou et al.\ 2002; Verde et al.\ 2003) also require a model for galaxy bias. Dekel \& Lahav (1999) show that nonlinearity and stochasticity in the bias relation can explain discrepancies between different methods of measuring parameters which assume a linear bias factor, such as measurements of $\beta=\Omega_{m}^{0.6}/b$ (Peacock et al.\ 2001; Hawkins et al.\ 2003).

In fact both theoretical approaches (Mo \& White 1996) and
simulations predict that bias may be non-linear and scale dependent,
at least on some (small) scales. Kauffmann et al.~(1997) find only weak scale dependence on large scales and a bias relation consistent with linear bias. Benson et al.~(2000) find that semi-analytic galaxies in a LCDM model could reproduce the APM correlation function given a scale dependent bias taking the form of an antibias of galaxies relative to matter on small scales. Somerville et al.~(2001) also use semi-analytic modelling to demonstrate that the physics of galaxy formation introduces a small scatter in the galaxy--mass relation; they find the mean bias to have only a weak dependence on scale for $r\lesssim 12h^{-1}$Mpc, (where the Hubble constant, $H_{0}=100$\,$h$\,km\,s$^{-1}$).

In principle the true mass distribution can be
directly measured from measurements of galaxy peculiar velocities
using e.g.~POTENT reconstruction (Dekel, Bertschinger \& Faber
1990). In practice accuracy is hard to achieve by such methods; the technique requires heavy smoothing since the error bars
per galaxy are large and the volumes surveyed up to the present are
relatively local. A useful probe is instead to compare the clustering of
different types of galaxy: if these cluster differently, at least one
type cannot exactly follow the mass distribution.

It has been known for some considerable time that galaxies of
different morphological type have different clustering
properties. Early-type galaxies, such as ellipticals or S0s, are
highly clustered, accounting for almost 90\% of galaxies in the cores
of rich clusters. This fraction drops off steeply, however, with
distance from the cluster cores and in the field 70\% of galaxies are
late-type galaxies: spirals and irregulars (Dressler 1980; Postman \&
Geller 1984). The level of fluctuations in each of the early- and
late-type density fields can also be compared using the correlation
functions or power spectra for the two sub-populations. This kind of
study is optimised for small separations ($\lesssim 10h^{-1}$Mpc) and
has generally revealed that the clustering amplitude of ellipticals is
greater than that of spirals by a factor of 1.3--1.5 (e.g.~Loveday et
al.~1995; Norberg et al.~2002a; Madgwick et al.~2003b). If both
density fields were perfectly correlated with the matter density field
this factor would be equivalent to the ratio between linear bias
parameters $(b_{E}(r)/b_{L}(r))^{2}$. There is also evidence that the
relative bias between sub-populations of galaxies is more complex than
the global galaxy bias. Measurements of the 2dFGRS bispectrum (Verde
\etal~2002) found no evidence for nonlinearity in the bias for 2dFGRS
galaxies. More recently however, Kayo \etal~(2004) find evidence for
relative bias being complex on weakly non-linear to non-linear scales
from a measurement of the redshift-space three-point correlation
function, as a function of galaxy colour and morphology, in the Sloan
Digital Sky Survey. Wild et al. (2004) have carried out a
counts-in-cells analysis using volume limited samples from the 2dFGRS,
and find evidence for non-linearity and stochastic effects.

A detailed framework for dealing with possible nonlinearities and
stochasticity in the bias relation is given by \citet{Dekel_1999},
based on the joint probability distribution of the galaxy and mass
densities $f(\delta_{g},\delta_{m})$. In an analogous manner we will
consider the joint probability distribution of the early- and
late-type galaxy density fields for magnitude limited samples in the
2dFGRS. This approach in large part follows the methods described in
Blanton (2000) for the Las Campanas Redshift Survey (LCRS), although
the geometry of the 2dFGRS is considerably more amenable to this kind
of study than that of the LCRS and allows us, for example, to examine
a large range of scales. 

This paper is organised into sections as follows. In Section~\ref{survey} we
summarize details of the 2dFGRS, the PCA-$\eta$ parameter and the
division into cells. We present a measurement of the variances of the counts in cells using the method of Efstathiou et al.~(1990), which we have corrected for integral constraint bias, in Section~\ref{variances}. In Section~\ref{dist_fns} we present an analysis of the one-point distribution of the counts in cells based on fits to a lognormal distribution function. In Section~\ref{rel_bias} we discuss the relative bias. We present the results of applying the `modified $\chi^{2}$' statistic of \citet{Tegmark_1999} to the joint counts in cells and then move on to describe the application of the maximum-likelihood technique of \citet{Blanton_2000} to constrain the relative bias between spectral types. We summarize our conclusions in Section~\ref{concl}.

\section{The 2\lowercase{d}F Galaxy Redshift Survey}\label{survey}

The 2dFGRS observations were carried out between May 1997 and April 2002 using the 2dF instrument: a multi-object spectrograph on the Anglo Australian Telescope. The main survey region consists of two broad strips, one in the South Galactic Pole region (SGP) covering approximately
$-37^\circ\negthinspace.5<\delta<-22^\circ\negthinspace.5$, $21^{\rm
h}40^{\rm m}<\alpha<3^{\rm h}40^{\rm m}$ and the other in the
direction of the North Galactic Pole (NGP), spanning
$-7^\circ\negthinspace.5<\delta<2^\circ\negthinspace.5$, $9^{\rm
h}50^{\rm m}<\alpha<14^{\rm h}50^{\rm m}$.  In addition there are a number of circular two-degree fields scattered randomly over the full extent of the low extinction regions of the southern APM galaxy survey. 

The parent catalogue for the survey was selected in the photometric
$\bj$\ band from a revised and extended version of the APM galaxy
survey (Maddox, Efstathiou \& Sutherland 1990a,b,c; 1996). The magnitude limit at the start of the survey was set at $\bj = 19.45$ but both the photometry of the input catalogue and the dust extinction map have since been revised and so there are small variations in magnitude limit as a function of position over the
sky. The effective median magnitude limit, over the area of the
survey, is $\bj \approx 19.3$ (Colless \etal~2001; Colless \etal~2003).

The completeness of the survey data varies according to the position
on the sky because of unobserved fields (mostly around the survey
edges), untargeted objects in observed fields (due to collision
constraints or broken fibres) and observed objects with poor spectra;
also there are drill-holes around bright stars. The variation in
completeness with angular position, $\btheta$, is fully described by
the completeness mask (Colless \etal~2001; Norberg \etal~2002b;
Colless \etal~2003). Note that since we use exclusively those galaxies
for which a principal component spectral type has been derived, we
require a slightly modified completeness mask from that describing the
completeness of the full survey, one which reflects the completeness of
galaxies with measured $\eta$-type (Norberg \etal~2002a).

We use the completed 2dFGRS data set which was released publicly at the end of June 2003 (Colless \etal~2003). This includes 221\,414 unique, reliable galaxy
redshifts (quality flag $\geqslant$ 3, Colless \etal~2001; Colless \etal~2003).  The random fields, which contain nearly 25\,000 reliable redshifts, are not included in this analysis. Throughout the paper we treat the NGP and SGP regions as independent data sets, which means we have two estimates for each of the statistics we derive. This approach functions both as a `reality check' for our error estimates and also gives an idea of the variation due to cosmic variance. 

\subsection{PCA classification of galaxy spectra}\label{PCA}

The spectral properties of 2dFGRS galaxies have been analysed and the
galaxies split into spectral type classes using a principal component
analysis (PCA) described by Madgwick et al. (2002). This technique
splits the galaxies on the basis of the characteristics of their
spectra which show the most variation across the sample, without using
any prior assumptions or template spectra. Madgwick \etal~(2002)
define a scalar parameter, $\eta$, which is a linear combination of
the first two principal components chosen to minimize instrumental
effects which make the determination of the continuum uncertain. In
effect, $\eta$ quantifies the relative strengths of emission and
absorption lines, and can be shown to be tightly correlated to the
equivalent width of \Halpha\ in particular, so a simple physical interpretation of $\eta$ is as a measure of the current star formation rate in a galaxy (Madgwick et al.~2003a).

The PCA classification makes use of the spectral information in the rest-frame wavelength range 3700\AA\ to 6650\AA, which includes all the major optical diagnostics between OII and \Halpha. The spectral coverage imposes a limit on the maximum redshift at which this analysis can be used of $z=0.2$. For galaxies with $z>0.15$ however, sky absorption bands contaminate the \Halpha\ line, which affects the stability of the classification. For this reason we restrict our analysis to galaxies with $z<0.15$ following Madgwick et al. (2002).

The distribution of \etapar\ for the 2dFGRS spectra is clearly bimodal
(see fig.~4 of Madgwick \etal~2002), with a shoulder at
$\eta=-1.4$. Madgwick \etal~(2002) divide galaxies into four spectral
type bins based on the shape of this distribution; the local minimum
at $\eta=-1.4$ is used to separate early and late types while the late
type `shoulder' is divided in two and also separated from the tail,
which will be dominated by particularly active galaxies such as
starbursts and AGN. Because of the effects of possible evolution in
the last two spectral type bins, discussed further in the next
section, we use only the spectral classes 1 \& 2 of Madgwick et al.\
(2002) in this paper, which we refer to as early and late type
respectively, and exclude the bluer classes 3 \& 4.

The important aspect of the PCA classification used here is that it
represents a coherent method for dividing a galaxy sample into classes
based on a diagnostic with a relatively clear physical interpretation,
i.e.~current star formation rate. Note that this specific
interpretation will not necessarily be the case for all galaxy samples
classified using a PCA method; comparing classifications based on a
PCA analysis between samples is in general nontrivial. In our case,
however, the fact that the PCA classification of the 2dFGRS is
dominated by \Halpha\ means that a classification of galaxies based on
$\eta$ is virtually the same as a classification based on the
equivalent width of \Halpha .

% The morphological classification of galaxies can be compared to the
% spectral classification defined by the $\eta$ parameter (Norberg et
% al.~2002; Madgwick, 2003). The median value of $\eta$ for each
% morphological type does correlate well with the classification bins
% described above; elliptical and S0 galaxies lie mainly in the
% early-type bin, and although there is considerable scatter in the
% $\eta$-classification of spirals, the median $\eta$ lies in the
% late-type bin. In practice we expect the correlation between $\eta$
% and morphological type to be rather tighter than this since these
% comparisons use nearby extended galaxies where the distribution of
% spectral types is expected to be distorted by aperture effects (see
% e.g. Kochanek, Pahre \& Falco 2000). Madgwick et al. (2002) show that
% aperture bias should be completely negligible beyond $z\sim 0.1$.

\subsection{Counts-in-cells}\label{CiC}

\begin{figure*}
\begin{minipage}[t]{\textwidth}
\begin{minipage}[t]{0.5\textwidth}
\includegraphics[angle=-90, width=0.97\textwidth]{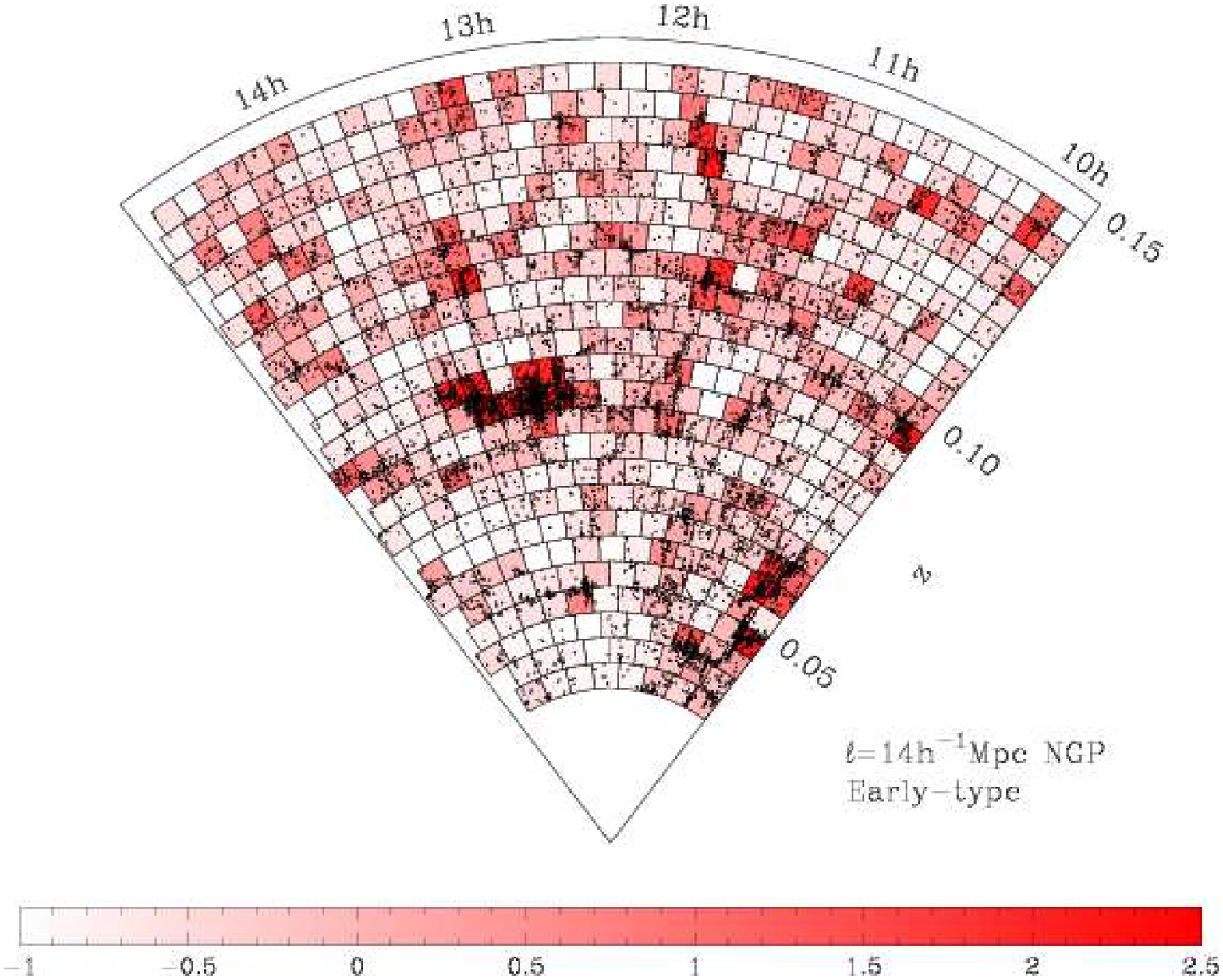}
\end{minipage}\hfill
\begin{minipage}[t]{0.5\textwidth}
\includegraphics[angle=-90, width=0.97\textwidth]{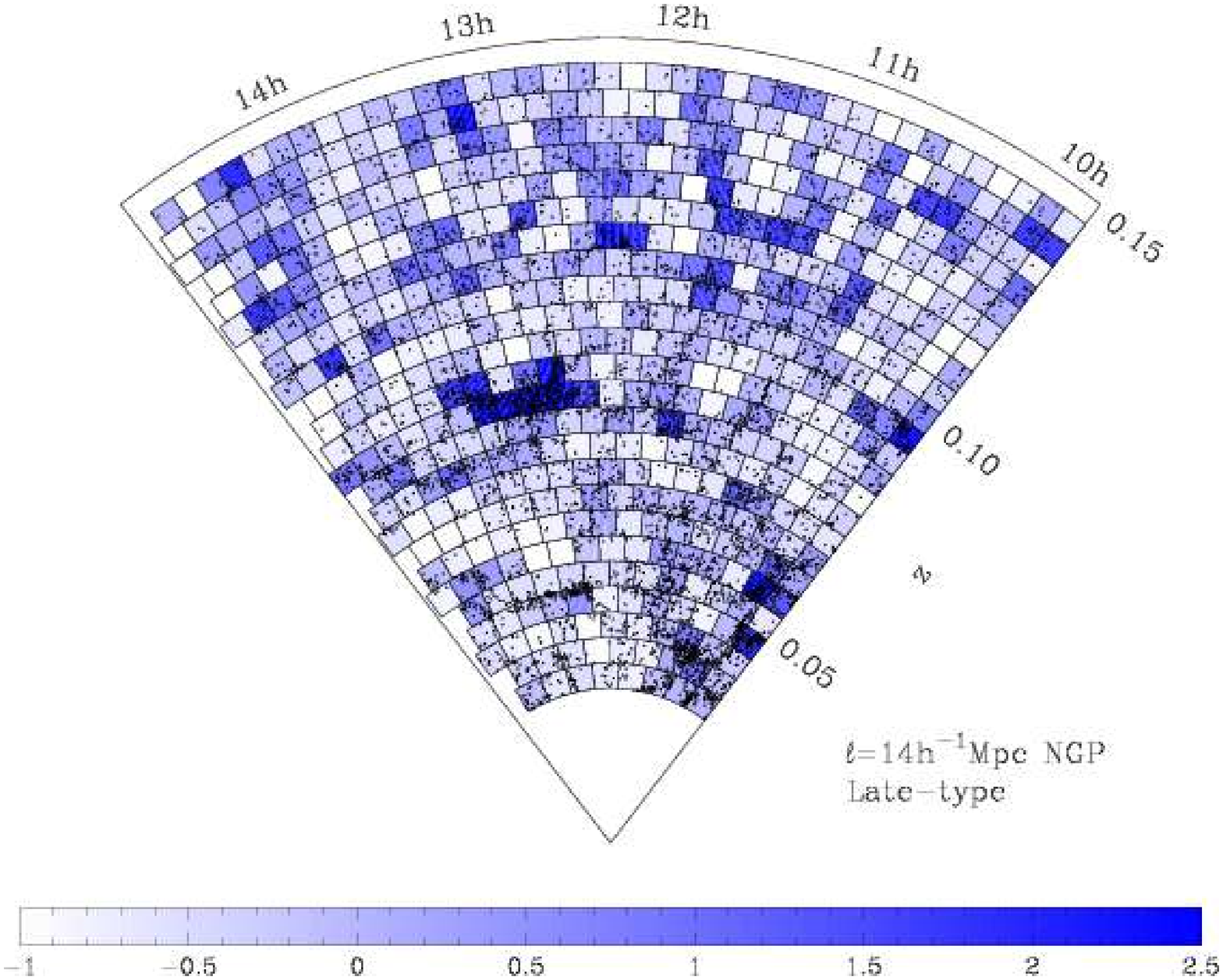}
\end{minipage}
\begin{minipage}[t]{0.5\textwidth}
\includegraphics[angle=-90, width=0.97\textwidth]{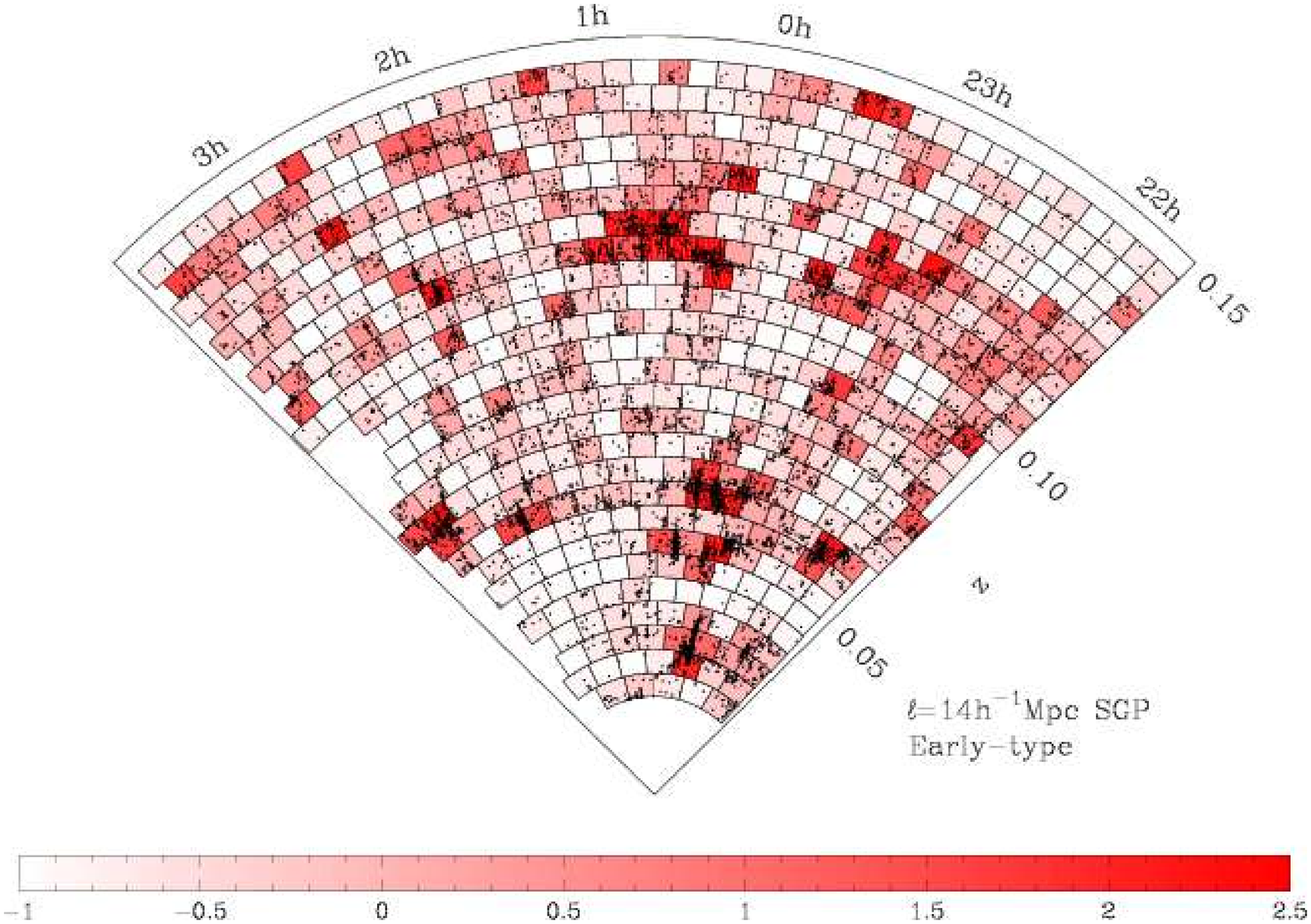}
\end{minipage}\hfill
\begin{minipage}[t]{0.5\textwidth}
\includegraphics[angle=-90, width=0.97\textwidth]{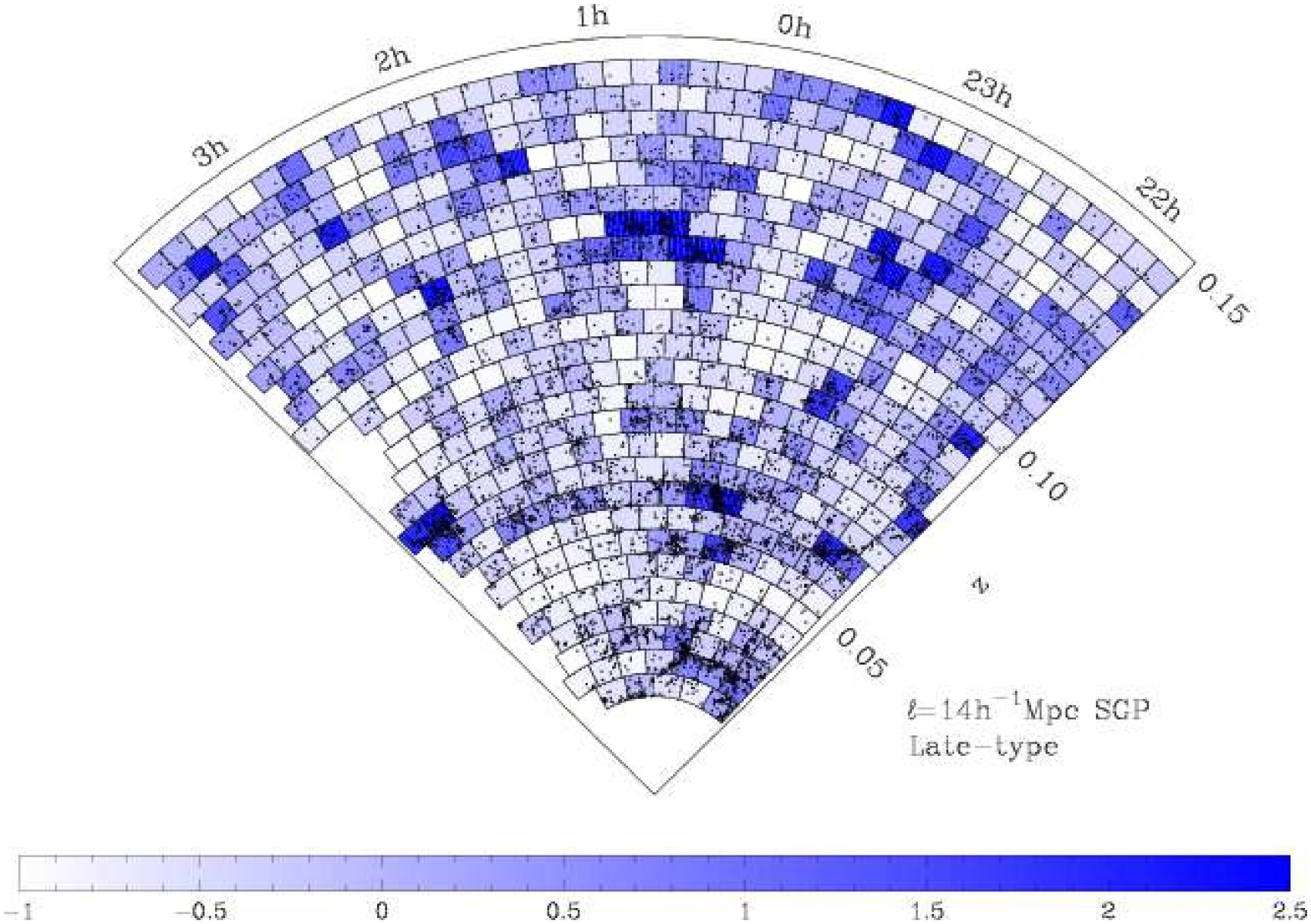}
\end{minipage}
\caption{An example of the division into cells for the NGP (top) and
  SGP (bottom) regions, with cells of side $\ell=14h^{-1}$Mpc; the cells
  spanning the central declination of each slice are plotted along
  with the galaxies of each type (early types in red \& on the left, late types
in blue \& on the right) in the cell. The colour scales
  indicate the estimated galaxy density contrast as defined in the
  text.}
\label{cuts}
\end{minipage}
\end{figure*}

The analysis presented in this paper is complementary to approaches to the
study of the relative bias based on correlation functions, in that we
compare the two density fields on a point by point
basis, rather than measuring overall clustering amplitudes. To
estimate the local galaxy density contrast for each type we use the
method of counts in cells. Analysis of the 2dFGRS counts in cells has already been used by Croton \etal\ (2004a,b) and Baugh \etal\ (2004) to constrain the higher-order correlation functions and the void probability function. Another respect in which this analysis is complementary to the correlation function approach is that it is optimised for much larger scales.

The method by which we divide the survey region into cells is
identical to that first used by \citet{Efstathiou_1990} for measuring
the variance of the counts in cells for a sparse-sampled redshift
survey of IRAS galaxies \citep{Rowan-Robinson_1990}. First the
surveyed region of space is divided into spherical shells of thickness
$\ell$ centred on the observer. Each shell is then subdivided into
approximately cubical cells using lines of constant right ascension
and declination, chosen for each radial distance and declination to
ensure that the sides of the cell of length $\approx\ell$, the shell
separation. In this paper we analyse cell divisions over a range in
$\ell$ from $\ell=7h^{-1}$Mpc to $\ell=42h^{-1}$Mpc. We assume a
concordance cosmology of $\Omega_{m}=0.3,
\Omega_{\Lambda}=0.7$. Effective scales corresponding to a given value
of $\ell$ can be computed using the approximate relationship between
the radius of a Gaussian sphere window, $R_{\rm G}$, the radius of a
spherical top-hat window, $R_{\rm T}$, and $\ell$, given in
Eq.~\ref{scales} (Peacock 1999).
\begin{equation}
\label{scales}
R_{\rm G} \sim \frac{R_{\rm T}}{\sqrt{5}} \sim \frac{\ell}{\sqrt{12}}.
\end{equation}

\begin{table*}
\centering
\begin{minipage}{130mm}
  \caption{Total number of cells and expected counts (presented as the 16\%--50\%--84\% percentiles of the distribution).}

\begin{tabular}{lcccccc}
\hline
 & \multicolumn{3}{c}{NGP} & \multicolumn{3}{c}{SGP} \\
$\ell$ ($h^{-1}$Mpc) & $N_{\rm cells}$ & $N_{{\rm exp},E}$ & $N_{{\rm exp},L}$ & $N_{\rm cells}$ & $N_{{\rm exp},E}$ & $N_{{\rm exp},L}$ \\
\hline
7.0  & 11056 & 2.0--3.2--5.7 & 1.2--2.2--4.7 & 14593 & 2.0--3.0--5.0 & 1.3--2.2--4.5 \\
8.75 & 5689 & 2.8--4.5--8.3 & 1.7--3.0--6.7 & 7543 & 2.7--4.1--6.9 & 1.8--3.0--6.2 \\
10.5 & 3170 & 4.0--6.5--11.7 & 2.5--4.4--9.5 & 4114 & 3.8--5.8--9.6 & 2.6--4.3--8.7 \\
12.25 & 2026 & 5.4--8.9--16.4 & 3.4--6.0--13.3 & 2567 & 5.3--7.8--13.1 & 3.6--5.9--12.0 \\
14.0 & 1198 & 8.4--13.1--23.3 & 5.3--8.8--18.9 & 1484 & 7.7--11.4--18.7 & 5.5--8.8--17.4 \\
17.5 & 620 & 13.7--22.8--40.5 & 8.7--15.7--32.6 & 729 & 12.7--18.3--30.0 & 9.2--14.4--28.3 \\
21.0 & 336 & 24.5--39.3--69.4 & 15.7--26.8--56.0 & 372 & 22.0--32.8--52.4 & 15.8--25.5--49.8 \\
24.5 & 187 & 41.3--64.6--117 & 25.8--43.3--95.6 & 205 & 36.0--49.2--80.3 & 26.6--38.6--75.9 \\
28.0 & 113 & 63.1--95.9--175 & 39.1--63.1--136 & 117 & 51.8--77.0--114 & 38.1--61.0--110 \\
31.5 & 80 & 98.1--162--252 & 61.8--109--197 & 125 & 86.3--122--179  & 62.6--99.5--171 \\
35.0 & 57 & 140--226--321 & 85.2--152--250 & 101 & 120--160--260 & 85.9--125--255 \\
38.5 & 45 & 183--278--390 & 112--184--301 & 77 & 143--175--305 & 104--138--296 \\
42.0 & 34 & 248--351--476 & 156--237--349 & 53 & 168--227--388 & 125--176--369 \\
\hline
\label{basic_stats}
\end{tabular}
\end{minipage}
\end{table*}

Any analysis of the galaxy counts in cells for a flux limited survey
must take into account the selection function, which quantifies
the probability that a galaxy with a given redshift, $z$, is included
in the survey. We define $M_{\rm min}(z,\btheta)$ and
$M_{\rm max}(z,\btheta)$ to be the minimum and maximum absolute magnitudes
visible at redshift $z$ given the magnitude limit of the survey
which, in the case of the 2dFGRS, varies with angular position $\btheta$
as described in Section~\ref{survey} and we take the luminosity function
$\Phi(M)$ to be normalized in that range. Then the selection function
can be written
\begin{equation}	
   \phi(z,\btheta) = \int_{M_{\rm min}(z,\btheta)}^{M_{\rm max}(z,\btheta)} dM~
\Phi(M)c_{z}(\btheta).
\end{equation}
The $c_{z}(\btheta)$ term describes the variation in completeness of the survey over the sampled area, for which we use the survey mask for $\eta$-typed galaxies described in Section~\ref{survey}. The completeness also depends slightly on apparent magnitude and, for the full survey, one can use the relation given in Colless \etal~(2001) to parameterise this variation:
\begin{equation}
c_{z}(\btheta,b_{\rm J})=\gamma\left\{1-\exp[b_{\rm J}-\mu(\btheta)]\right\},
\end{equation}
where $\gamma$ is set at $0.99$ and $\mu(\btheta)$ can be set by the requirement that
\begin{equation}
\left\langle c_{z}(\btheta,b_{\rm J}) \right\rangle = c_{z}(\btheta),
\end{equation}
where the average is over all apparent magnitudes, $b_{\rm J}$.

Unfortunately, the $\mu(\btheta)$ mask is undefined for $\eta$-typed galaxies. To generate such a mask one would need to assume a form for the number counts as a function of $\eta$, as well as making an assumption for the variation of completeness with limiting redshift, $z_{\rm max}$, since the $\eta$-parameter is only defined for $z\leq0.15$. Given the large number of assumptions which would be necessary we have not included a correction for the effect of apparent magnitude on completeness in this study; the effect is in any case a small one, particularly in a flux limited catalogue.

Once we have a knowledge of the selection function we can define the
expected number of galaxies in each cell $i$ by
\begin{equation}
N_{{\rm exp},i} = \int_{V_{i}} dV_{i}~\phi(z,\btheta)
\end{equation}
Where the integral is over the volume of the cell $i$. 

We use expected counts in cells ($N_{E,{\rm exp},i}$ and $N_{L,{\rm
exp},i}$ for early and late types respectively), which we obtain by
integrating the \etapar-type dependent luminosity functions of
\citet{Madgwick_2002} using the average magnitude limit over the
surface of each cell, (Colless \etal~2001; Colless \etal~2003) and
corrected for $c_{z}(\btheta)$ as described above. We reject from the
analysis cells for which the average completeness over the cell is
less than 70\%. We also renormalize the expected counts to ensure that
$\left\langle N/N_{\rm exp}\right\rangle =1$ in order to correct for
possible errors in the normalization of the luminosity functions. We
find that this choice of renormalization gave the most stable results
although the exclusion of empty cells from the renormalization step
was necessary to ensure stability (see Section~\ref{empty_cells}). In
practice the details of this renormalization step do not significantly
affect the best fitting parameters for the variance or relative bias,
but they do change the KS test probabilities for our models.

This approach to dealing with the selection function for subsets of
the data, such as the division into early and late-type galaxies, is
vulnerable to systematic errors in the selection function. An example
of such an effect would be a surface-brightness term in the selection
function for $\eta$-typed galaxies, which one would expect would
affect early- and late-type galaxies differently. Any systematic error
in calculating the expected number of counts in cells for the early-
or late-type galaxies will bias results for the nonlinearity and
stochasticity of the bias function. Such an effect was noted in
Blanton (2000), who found that his results were sensitive to excluding
the low redshift part of his sample. Madgwick \etal~(2002) detect a
large overabundance of spectral types 3 and 4 beyond $z=0.11$,
relative to the predicted $n(z)$ based on the luminosity functions for
these spectral types. Although such an observation could be due to
aperture effects, a more plausible explanation is the presence of
evolution for these spectral type bins. In principle, such evolution
could be modelled in our analysis, and we could derive accurate
$N_{\rm exp}$ for all late-type galaxies. If there is evolution in
these spectral types however, we may expect that the relative bias
could also be evolving over the redshifts used in this analysis. For
this reason we have used only $\eta$-types 1 \& 2 in the analysis.

Fig.~ \ref{cuts} shows the division into cells for
$\ell=14h^{-1}$Mpc. The galaxies in each cell of the respective
spectral type are shown overlaid on a colour scale indicating the
estimated galaxy density contrast in that cell. The intermediate
density contrasts are much less prevalent in the early-type density
field, implying that the contrast between clusters and voids is increased,
as one would expect. Table~\ref{basic_stats} shows the number of cells
in each cell division together with the median and 16\% and 84\%
percentiles of the distribution of expected counts.

Wild et al. (2004) have also analysed colour selected volume limited
samples from the 2dFGRS. Unfortunately, luminosity functions for
different colour selections from the 2dFGRS have not yet been
measured, and so we cannot define the selection functions required to
carry out our analyses on colour selected magnitude limited samples.

\section{The variance of the counts in cells}\label{variances}

One of the most fundamental statistics accessible from the counts in cells is the variance of the counts. This is directly related to the galaxy autocorrelation function, being equal to the volume average of the correlation function once the contribution of discreteness noise is removed. In later sections we will fit a parametric model to the one-point distribution of the counts in cells as the first stage in a maximum likelihood approach to fitting for the relative bias; we would expect that an accurate model for the PDF will reproduce the variance of the counts as measured in this section. Furthermore, by comparing the variance between spectral types, we can obtain an estimate of the linear relative bias parameter.

\subsection{Predictions for the variance}\label{predictions}

The real-space and redshift-space correlation functions for the full set of 2dFGRS galaxies have been accurately measured by Hawkins \etal~(2003). We have used these results to obtain predictions for the variance of the counts in cells using a number of approaches outlined below.

If we assume that the real-space correlation function is well described by a power-law of the form 
\[
\xi(r)=\left(\frac{r}{r_{0}}\right)^{-\gamma },
\]
then we can use the following form for the power spectrum expressed as the variance per $\ln k$:
\begin{equation}
\label{delt_k}
\Delta^{2}(k)=\frac{2}{\pi}(kr_{0})^{\gamma}\Gamma(2-\gamma)\sin\frac{(2-\gamma)\pi}{2}.
\end{equation}
For the case of Gaussian spheres of radius ${\rm R}_{\rm G}$, the variance in spheres, $\sigma^{2}$, is equal to the value of $\Delta^{2}(k)$ at 
\begin{equation}
\label{k_value}
k=\left[\frac{1}{2}\left(\frac{\gamma-2}{2}\right)\!\right]^{1/\gamma}{\rm R}_{\rm G}^{-1},
\end{equation}
but this is also a good approximation to the variance in cubical cells if we take ${\rm R}_{\rm G}=\ell/\sqrt{12}$, as described in Section~\ref{CiC}.

\begin{figure*}
\begin{minipage}[t]{\textwidth}
\begin{center}
\includegraphics[angle=-90, width=0.7\textwidth]{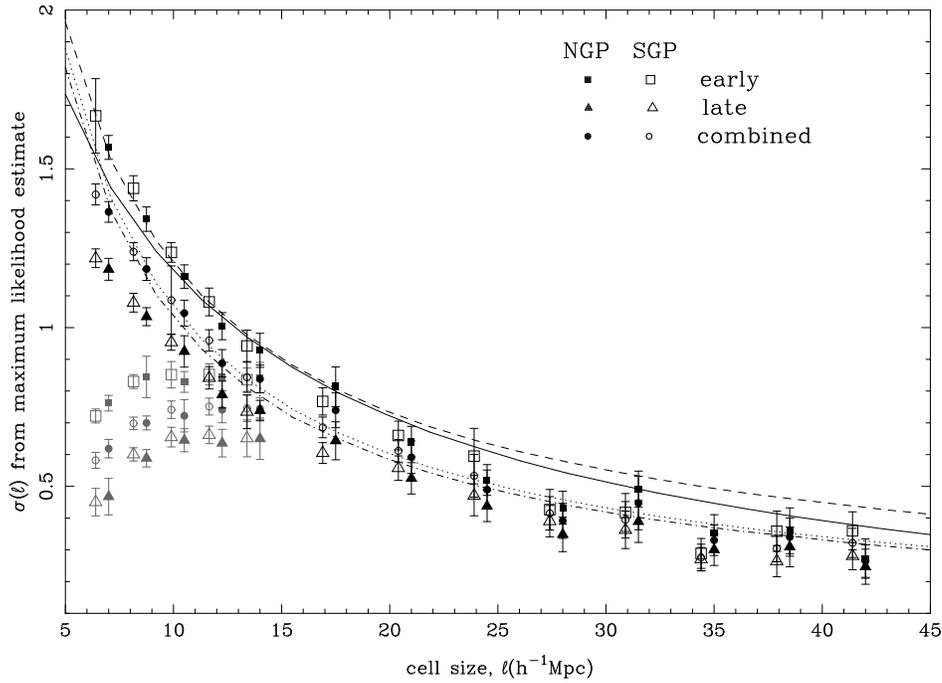}
\end{center}
\caption{$\sigma$ as a function of cell size, $\ell$, as measured by the maximum likelihood technique of Efstathiou et al.~(1990). Filled symbols are for the NGP region,
  open symbols are SGP. The results shown are fits to the early-type
  galaxies (squares), late types (triangles), and to both types
  combined (circles). Predicted values are overlaid, calculated using a power-law form for $\Delta^{2}(k)$ (Eq.~\protect\ref{delt_k}; dotted line); from the integral over the power-law fit to $\xi(r)$ (dash-dot line), from the integral over the power-law fit to the redshift-space correlation function, $\xi(s)$ (dashed line), and from the integral over the interpolated data table for $\xi(s)$ (solid line). (The points for the SGP are offset by a small
  amount for clarity). Points in grey are the measured values when empty cells are removed from the analysis; for $\ell>14h^{-1}$Mpc this has no effect on the variance measurements. The error bars are derived from Monte-Carlo realizations of lognormal models as described in the text.
}
\label{efstat_sig_plot}
\end{minipage}
\end{figure*}

More accurate predictions can be obtained by directly integrating the correlation function over a cubical volume of side $\ell$ to calculate the variance in cells,
\begin{equation}
\sigma^{2}=\frac{1}{V^{2}}\int_{V}dV_{1}~dV_{2}~\xi(r).
\end{equation}
We have used this method to calculate predictions based on both the
real-space and redshift-space correlation functions of Hawkins
\etal~(2003). The variance obtained from a volume average of the best-fitting power-law form for the real-space correlation function (Hawkins
\etal~2003) is almost identical to that using the scaling relation
(Eqs.~\ref{delt_k} \& \ref{k_value}) as one might expect. We would not
expect our measured variances to match the real-space predictions as
we have not made any corrections for redshift-space distortions. The
variances predicted using the redshift-space $\xi(s)$ from Hawkins
\etal~(2003) are significantly higher. We have calculated
redshift-space variance estimates using both a power-law approximation
for $\xi(s)$ and a direct interpolation from the data of Hawkins
\etal~(2003), since $\xi(s)$ is not well approximated by a power
law. The estimated variance from the interpolated $\xi(s)$ data
(solid lines in Figs~\ref{efstat_sig_plot}, \ref{efstat_sig_plot_c} \&
\ref{sig_plot}) is likely to give the most accurate prediction for the
variance of the counts in cells for the combination of spectral types
1 \& 2.

\subsection{Measuring the variance from the counts in cells}\label{ef_test}

A maximum-likelihood technique for calculating the variance of counts in cells is presented by \citet{Efstathiou_1990}. In each redshift shell in our cell division we can compute the statistic
\begin{equation}
\label{S}
S=\frac{1}{M-1}\sum_{i}(N_{i}-\bar{N})^{2}-\bar{N},
\end{equation}
where the sum extends over the M cells in the shell, and $\bar{N}$ is the mean cell count; $N_{i}$ are the observed counts in cell $i$. Note that this technique is based only on the measured counts in cells and does not use the expected counts, $N_{\rm exp}$, calculated in Section~\ref{CiC}. The expectation value for $S$ is
\begin{equation}
\left\langle S\right\rangle=n^{2}V^{2}\sigma^{2}=\bar{N}^{2}\sigma^{2},
\end{equation}
where $n$ is the mean number density and $V$ is the volume of the cells. The variance of $S$ for the case where the underlying density fluctuations are Gaussian is given by
\begin{equation}
\label{SVar}
{\rm Var}(S)=\frac{2n^{2}V^{2}(1+\sigma^{2})+4n^{3}V^{3}\sigma^{2}+2n^{4}V^{4}\sigma^{4}}{M}.
\end{equation}

Clearly this variance will be underestimated since we have made two key assumptions which are not strictly correct; namely that the underlying fluctuations are Gaussian and that the cells are independent. The effect of correlations between cells was addressed by Broadhurst, Taylor \& Peacock (1995). They show that for all but adjacent cells the covariance in the cell counts will be negligible and, within the accuracy to which $\sigma^{2}$ can be calculated, the error in treating even adjacent cells as independent is unimportant. A far more serious concern is with the assumption of Gaussian fluctuations; even on the relatively large scales of this analysis this assumption is far from valid. By using the variance estimator of Eq.~\ref{S} on Monte-Carlo realizations of lognormal fields we find that the variance of $S$ is many times larger than expected from Eq.~\ref{SVar}.

This method must be modified to deal with completeness variations in the 2dFGRS, as quantified by the survey mask. \citet{Efstathiou_1990} gives the following modified estimator
for $S$ for the case where cells are incomplete due to the survey
mask:
\begin{eqnarray}\label{mod_S}
S=&& \hspace{-0.73cm}A/B, \nonumber\\ 
{\rm where} \nonumber \\
A=&&\hspace{-0.73cm}\left(\frac{\ell^{3}}{M}\sum_{i}\frac{N_{i}}{V_{i}}\right)^{2}\hspace{-0.15cm}\left[\sum_{i}\left(N_{i}-V_{i}\frac{n_{1}}{v_{1}}\right)^{2}\hspace{-0.1cm}-n_{1}\left(1-\frac{v_{2}}{v_{1}^{2}}\right)\right] \nonumber \\
%&&\hspace{-0.6cm}\left. -n_{1}(1-v_{2}/v_{1}^{2})\right\}, \nonumber\\
B=&&\hspace{-0.73cm}\left(\frac{n_{1}}{v_{1}}\right)^{2}\left[v_{2}-2\frac{v_{3}}{v_{1}}+\frac{v_{2}^{2}}{v_{1}^{2}}\right],
\end{eqnarray}
where for convenience we have used the quantities $n_{1},v_{1},v_{2},v_{3}$, defined as:
\begin{equation}
n_{1}=\sum_{i}N_{i},~v_{1}=\sum_{i}V_{i},~v_{2}=\sum_{i}V^{2}_{i},~v_{3}=\sum_{i}V^{3}_{i},
\end{equation}
where $V_{i}$ is the usable volume of cell $i$. This correction is only valid if only a small fraction of a cell is excluded by the mask, so for this test we reject all cells where the fraction of the cell of low completeness ($<70$\%) is less than 30\%. We also upweight the counts in cells to compensate for incompleteness, based on the survey mask.

For each shell, $j$, we define a likelihood,
\begin{equation}
\label{ef_likelihood}
L_{j}(\sigma)=\frac{1}{\sqrt{2\pi{\rm Var}(S_{j})}}\exp\left[\frac{-(S_{j}-n_{j}^{2}V^{2}\sigma^{2})^{2}}{2{\rm Var}(S_{j})}\right],
\end{equation}
which is calculated using Eq.~\ref{mod_S} as an estimator for $n_{j}^{2}V^{2}\sigma^{2}$. We then minimise with respect to $\sigma$ the quantity 
\begin{equation}
\mathcal{L} \equiv -2\sum_{j} \ln L_{j}, \label{L_define} 
\end{equation}
where the sum is over all shells. Note that this differs by a factor 2
compared to $\mathcal{L}$ defined by Wild et al.

As we discussed above the variance on the estimator, ${\rm Var}(S)$, is in fact badly underestimated by Eq.~\ref{SVar} for realistic non-Gaussian density fluctuations. Although in practice the procedure adopted by \citet{Efstathiou_1990} of deriving errors in $\sigma$ from the likelihood function will not underestimate errors as dramatically as this, since the variation between shells will contribute to the error estimate for $\sigma$, we have instead used Monte-Carlo realizations of lognormal fields at the appropriate variance to derive more realistic error bars. Even though these errors are model dependent, the density fluctuations are much more closely approximated by a lognormal model than the Gaussian assumption of Eq.~\ref{SVar}.

\subsubsection{Cell variances and estimation bias}

The variances of the counts in cells calculated using the maximum likelihood approach are shown in Fig.~\ref{efstat_sig_plot}. Since we have upweighted the counts in cells to compensate for completeness effects we should exclude cells with zero counts from the analysis. This inevitably leads to an underestimated variance, particularly on small scales where the empty cells become significant ($\ell\lesssim15h^{-1}$Mpc; grey points in Fig.~\ref{efstat_sig_plot}). Below this scale we have nevertheless calculated the variance considering empty cells as being genuinely empty (black points in Fig.~\ref{efstat_sig_plot}). The difference between these two sets of points provides a means of estimating the magnitude of any bias introduced by excluding empty cells which will be relevant in the following section.

 The variances of the counts in cells for all $\eta$-typed galaxies seem to be consistent with the predicted variances from the real-space correlation function, $\xi(r)$, which is an interesting result since the calculation is carried out in redshift space, with no corrections for redshift-space distortions. This is due to an estimation bias leading to an underestimate of the true redshift-space variance in cells; the fact that this results in a variance consistent with the real-space variance would appear to be a coincidence.

The estimation bias in question is a finite volume effect exactly equivalent to the integral constraint, and is discussed in detail in Hui \& Gazta\~naga (1999). It becomes relevant here because we are using radial shells which have a rather small volume because of the geometry of the 2dFGRS slabs. Hence the individual shell contributions, $S_{j}$ in Eq.~\ref{ef_likelihood}, can be significantly biased. 

Hui \& Gazta\~naga give a useful analytical approximation for the magnitude of the integral constraint bias in the variance. We express the expected value of our variance estimator as
\begin{equation}
\left\langle\hat{\sigma}^{2}\right\rangle\equiv\sigma^{2}\left(1+\frac{\Delta_{\sigma^{2}}}{\sigma^{2}}\right),
\end{equation}
where $\Delta_{\sigma^{2}}/\sigma^{2}$ is the fractional bias in $\sigma^{2}$. The fractional bias can be approximated by the expression:
\begin{equation}
\frac{\Delta_{\sigma^{2}}}{\sigma^{2}}=-\frac{\sigma^{2}_{V}}{\sigma^{2}} + (3-2c_{12})\sigma^{2}_{V},
\end{equation}
where $\sigma^{2}_{V}$ is the two-point correlation function averaged over the whole volume in question, which in this case is the volume of a shell, and $c_{12}$ is a coefficient derived from the hierarchical relation:
\begin{equation}
\left\langle\delta^{m}_{i}\delta^{m'}_{j}\right\rangle_{c}=c_{mm'}\bar{\xi}^{m+m'-2}_{2}\xi_{2}(i,j),
\end{equation}
where $\left\langle\delta^{m}_{i}\delta^{m'}_{j}\right\rangle_{c}$ is the connected cosmic $m + m'$-point function, neglecting Poisson terms, with at most two differing indices. We have used the perturbative value for $c_{12}$ from Bernardeau (1994):
\begin{equation}
c_{12}=68/21+\gamma/3.
\end{equation}

\begin{figure}
\begin{center}
\includegraphics[angle=-90, width=0.45\textwidth]{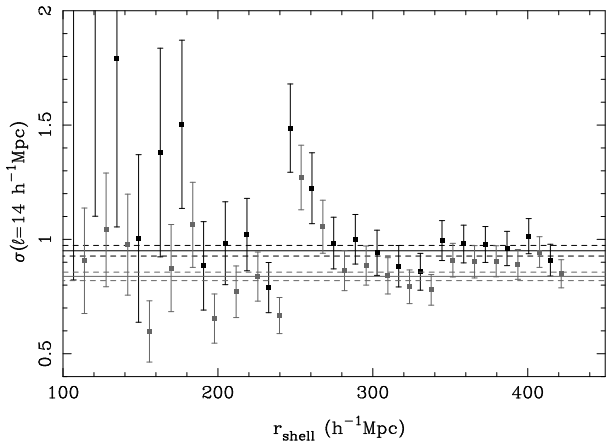}
\end{center}
\caption{An example of the estimated cell variance in shells compared to the maximum likelihood value and its associated 1-$\sigma$ error shown in the horizontal black solid and dashed lines. This example is for the $\ell=14h^{-1}$Mpc cell division of all $\eta$-typed galaxies in the NGP region. The points in grey (offset for clarity) show the same plot before the effects of the integral constraint bias have been corrected for and the grey horizontal solid and dashed lines show the uncorrected maximum likelihood estimate with its associated 1-$\sigma$ error.}
\label{eg_sig_plot}
\end{figure}

\begin{figure*}
\begin{minipage}{\textwidth}
\begin{center}
\includegraphics[angle=-90, width=0.7\textwidth]{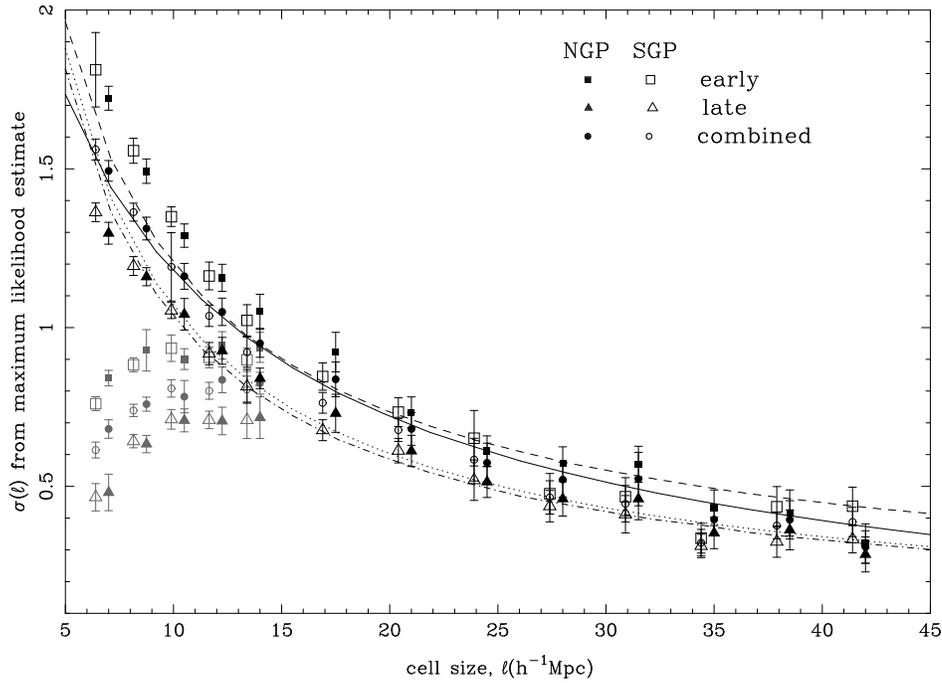}
\end{center}
\caption{The same as Fig.~\protect\ref{efstat_sig_plot} but using the approximation of Hui \& Gazta\~naga to correct for the integral constraint bias}
\label{efstat_sig_plot_c}
\end{minipage}
\end{figure*}

Using this approximation we can correct the $S_{j}$ and ${\rm Var}(S_{j})$ in Eq.~\ref{ef_likelihood} to obtain a bias-corrected estimate for $\sigma$. In Fig.~\ref{eg_sig_plot} we show an example of the estimated variances in shells for both the original Efstathiou estimator (Eq.~\ref{mod_S}; grey points and horizontal lines) and the bias-corrected version (black points and horizontal lines), for a single scale $\ell=14h^{-1}$Mpc. The correction for integral constraint bias shifts the maximum-likelihood variance estimator such that our results for the variance of the counts in cells for all $\eta$-typed galaxies are now consistent with the predicted values from the redshift-space correlation function as shown in Fig.~\ref{efstat_sig_plot_c}. The bias correction also increases the errors on the individual shell variance measurements. 

The effect of large-scale structure can clearly be seen in Fig.~\ref{eg_sig_plot}; in particular the noticeable spike around $r=250h^{-1}$Mpc corresponds to the prominent group of large clusters in the NGP region at around $z=0.09$. This is the NGP `hotspot' observed by Baugh \etal~(2004). Removing the shells around this value of $r$ does not alter our results appreciably relative to the magnitude of the estimated errors. 

\section{One-point distribution functions}\label{dist_fns}

Given a model for the one-point density distribution function, we can
model the one-point distribution of galaxy counts as being
equivalent to a convolution of the density field with Poisson
fluctuations of intensity $\lambda=N_{\rm exp}(1+\delta)$, i.e.
\begin{equation}
P(N) = \int
d\delta~\frac{N^{N}_{\rm exp}(1+\delta)^{N}}{N!}e^{-N_{\rm exp}(1+\delta)}f(\delta)
\label{blanton_eq1}
\end{equation}

We can then use Eq.~\ref{blanton_eq1} as the basis for a maximum-likelihood method to determine the parameters of the best fit
model for $f(\delta)$. A number of models have been proposed for $f(\delta)$ including the lognormal distribution (Coles \& Jones 1991), negative binomial model (Fry 1986; Carruthers 1991; Gazta\~naga \& Yokoyama 1993; Bouchet \etal~1993), Edgeworth expansion around the Gaussian distribution (Juszkiewicz \etal~1995) and Edgeworth expansion around the lognormal model -- the skewed lognormal approximation of Colombi (1994). Ueda \& Yokoyama (1996) consider fits of all of the above models to the counts in cells of a low density CDM N-body simulation. They find that the most satisfactory fit is given by the skewed lognormal model but unfortunately it is not positive definite, making it unsuitable for the maximum likelihood fitting procedure outlined below. The lognormal model is a satisfactory fit to the data on most scales except for the highly non-linear regime; it is also the most mathematically convenient since the version given below is already normalized in the interval $-1 \leq \delta < \infty$ and ensures that $\left\langle \delta \right\rangle = 0$. In later sections we will use the best fit $f(\delta)$ to determine the parameters for a number of models for the relative bias, following the example of \citet{Blanton_2000}.

\subsection{Lognormal model fitting}

\begin{table*}
\centering	
 \begin{minipage}{87mm}
  \caption{Best-fitting lognormal model parameters and KS-test probabilities for the lognormal model fit. The parameters for low $\ell$ cell divisions are derived from the data excluding empty cells and corrected for the resultant bias as described in the text.}
\begin{tabular}{ccccccc}
\hline
$\ell$ ($h^{-1}$Mpc) & $\sigma_{{\rm LN},E}$ & $P_{\rm KS}$ & $\sigma_{{\rm LN},L}$ & $P_{\rm KS}$ & $\sigma_{\rm LN, all}$ & $P_{\rm KS}$  \\
\hline
7.0  & 1.20  & 6e-14   & 1.05  & 7e-55  & 1.12  & 2e-99   \\
8.75 & 1.12  & 0.001   & 0.94  & 3e-18  & 1.03  & 4e-47   \\
10.5 & 1.05  & 0.229   & 0.88  & 0.002  & 0.97  & 5e-15   \\
12.2 & 0.99  & 0.268   & 0.81  & 0.134  & 0.90  & 0.001    \\
14.0 & 0.95  & 0.105   & 0.77  & 0.183  & 0.87  & 0.015      \\
17.5 & 0.84  & 0.292   & 0.67  & 0.442  & 0.77  & 0.212      \\
21.0 & 0.78  & 0.670   & 0.61  & 0.193  & 0.70  & 0.332      \\
24.5 & 0.68  & 0.291   & 0.53  & 0.710  & 0.61  & 0.414      \\
28.0 & 0.59  & 0.992   & 0.45  & 0.695  & 0.53  & 0.987      \\
31.5 & 0.57  & 0.829   & 0.44  & 0.988  & 0.51  & 0.962      \\
35.0 & 0.53  & 0.553   & 0.41  & 0.625  & 0.48  & 0.817      \\
38.5 & 0.48  & 0.195   & 0.39  & 0.308  & 0.44  & 0.965      \\
42.0 & 0.42  & 0.850   & 0.32  & 0.925  & 0.37  & 0.825      \\

\hline
\label{KS_probs}
\end{tabular}
\end{minipage}
\end{table*}

 Provided that the model for $f(\delta)$ that we choose is a reasonably accurate approximation, the actual choice of model should not affect the conclusions we infer for the relative bias. We therefore use exclusively the lognormal model, given in Eq.~
\ref{lognormal}, where $x =\ln (1+\delta) + \sigma_{\rm LN}^{2}/2$,
\begin{equation}
\label{lognormal}
f(\delta)d\delta =
\frac{d\delta}{\sigma_{\rm LN}\sqrt{2\pi}(1+\delta)}\exp\left(-\frac{x^{2}}{2\sigma_{\rm LN}^{2}}\right).
\end{equation}

It should be noted that we have here followed the notation of Coles \&
Jones (1991) in that $\sigma_{\rm LN}^{2}$ is the variance of the Gaussian
model from which the lognormal is derived by transformation -- note that Wild \etal\ (in preparation) use $\omega^{2}$ for the same parameter. The variance for the lognormal model is given by
\begin{equation}
\label{var_LN}
\left\langle\delta^{2}\right\rangle=\exp(\sigma_{\rm LN}^{2})-1.
\end{equation}
To ensure that the results of this section can be easily compared to those measuring the variance in the previous section we have used the above relation to transform our variances from $\sigma_{\rm LN}$ to $\sigma(\ell)\equiv\sqrt{\left\langle\delta^{2}\right\rangle}$.

Using this model, we define a likelihood for each cell $i$, as 
\[
L_{i} \equiv P(N_{i}|\sigma_{\rm LN}),
\]
in which the probability of observing $N_{i}$ galaxies in cell $i$ is determined by Eq.~
\ref{blanton_eq1}. We then find $\sigma_{\rm LN}$ for the best fit
lognormal model by minimizing with respect to $\sigma_{\rm LN}$ the
quantity
\begin{equation}
\mathcal{L} \equiv -2\sum_{i} \ln L_{i},
\label{curlyL}
\end{equation}
where the sum is over all cells, as defined previously in
Eq.~\ref{L_define}.  Again, this differs by a factor 2 compared to
$\mathcal{L}$ defined by Wild et al.

\begin{figure*}
\begin{minipage}[t]{\textwidth}
\begin{minipage}[t]{0.33\textwidth}
\includegraphics[angle=-90, width=0.97\textwidth]{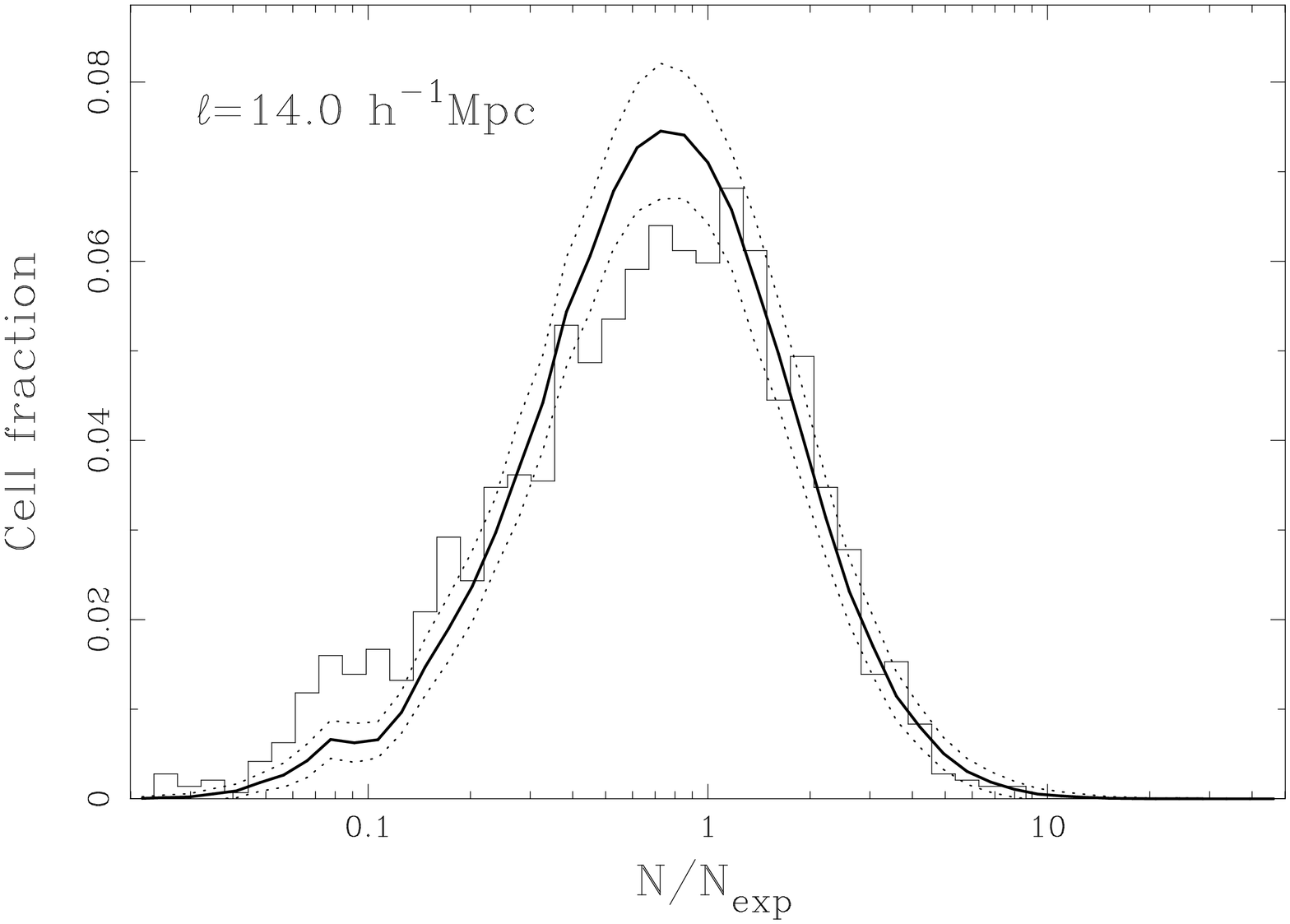}
\end{minipage}\hfill
\begin{minipage}[t]{0.33\textwidth}
\includegraphics[angle=-90, width=0.97\textwidth]{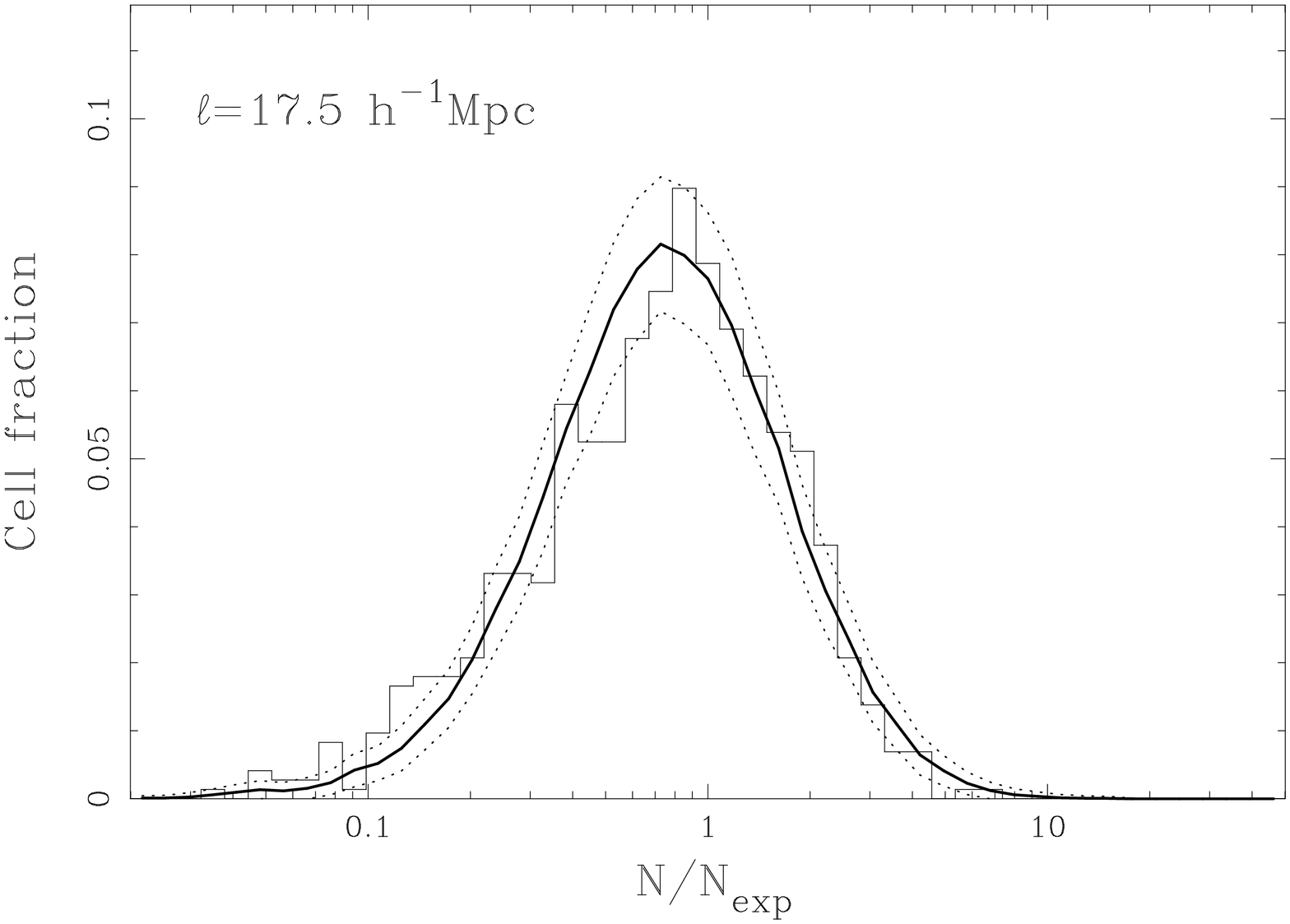}
\end{minipage}
\begin{minipage}[t]{0.33\textwidth}
\includegraphics[angle=-90, width=0.97\textwidth]{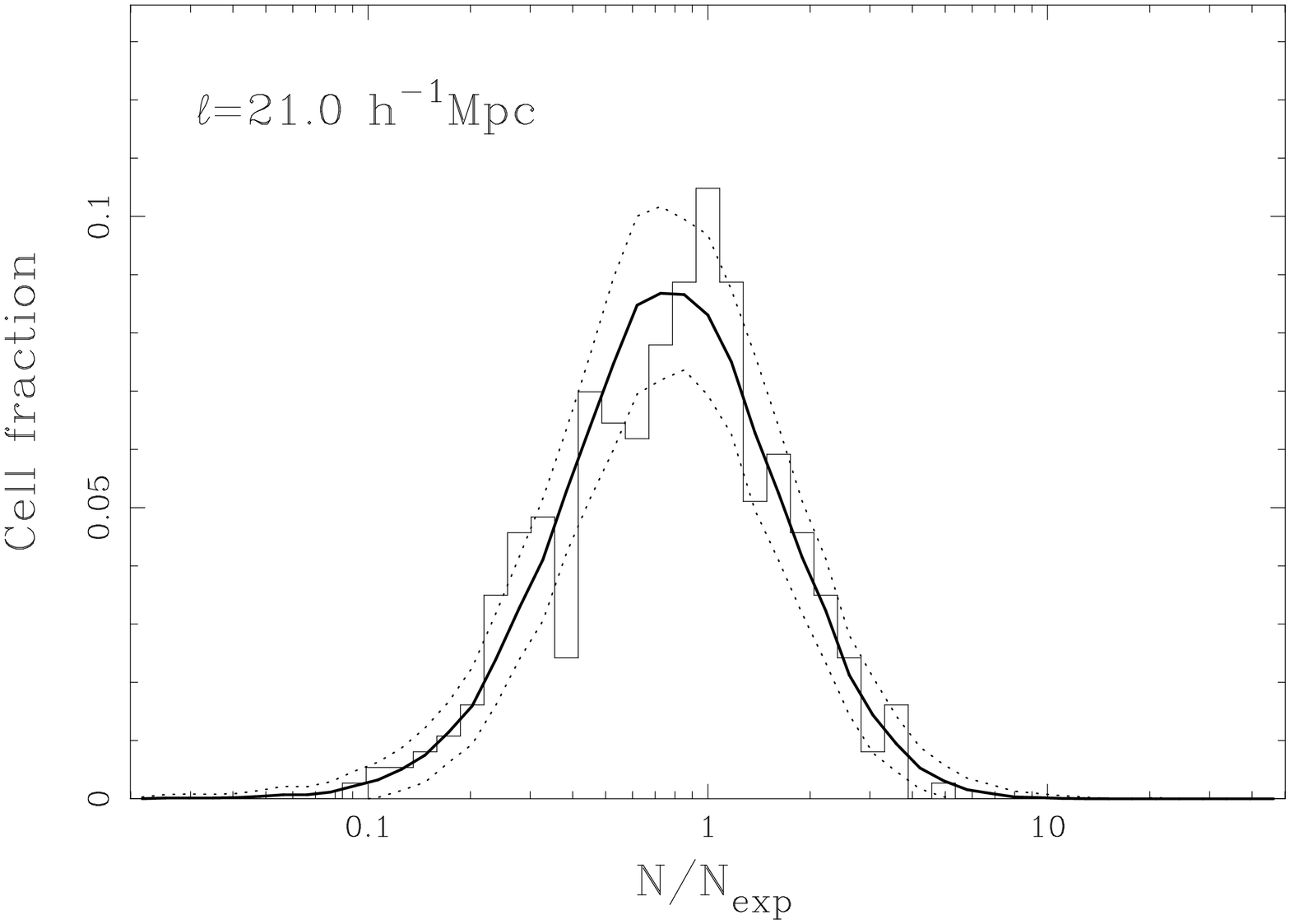}
\end{minipage}
\begin{minipage}[t]{0.33\textwidth}
\includegraphics[angle=-90, width=0.97\textwidth]{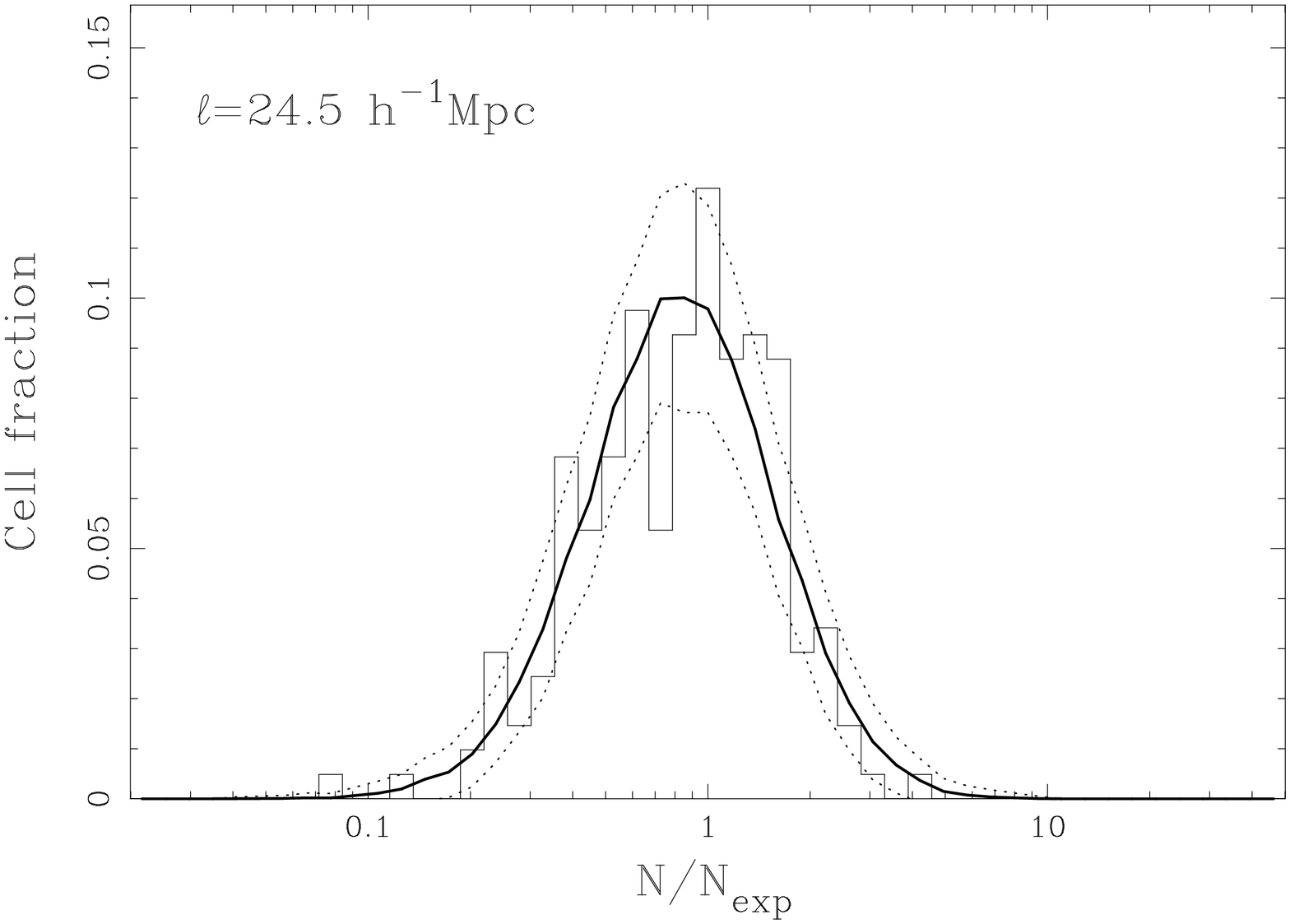}
\end{minipage}\hfill
\begin{minipage}[t]{0.33\textwidth}
\includegraphics[angle=-90, width=0.97\textwidth]{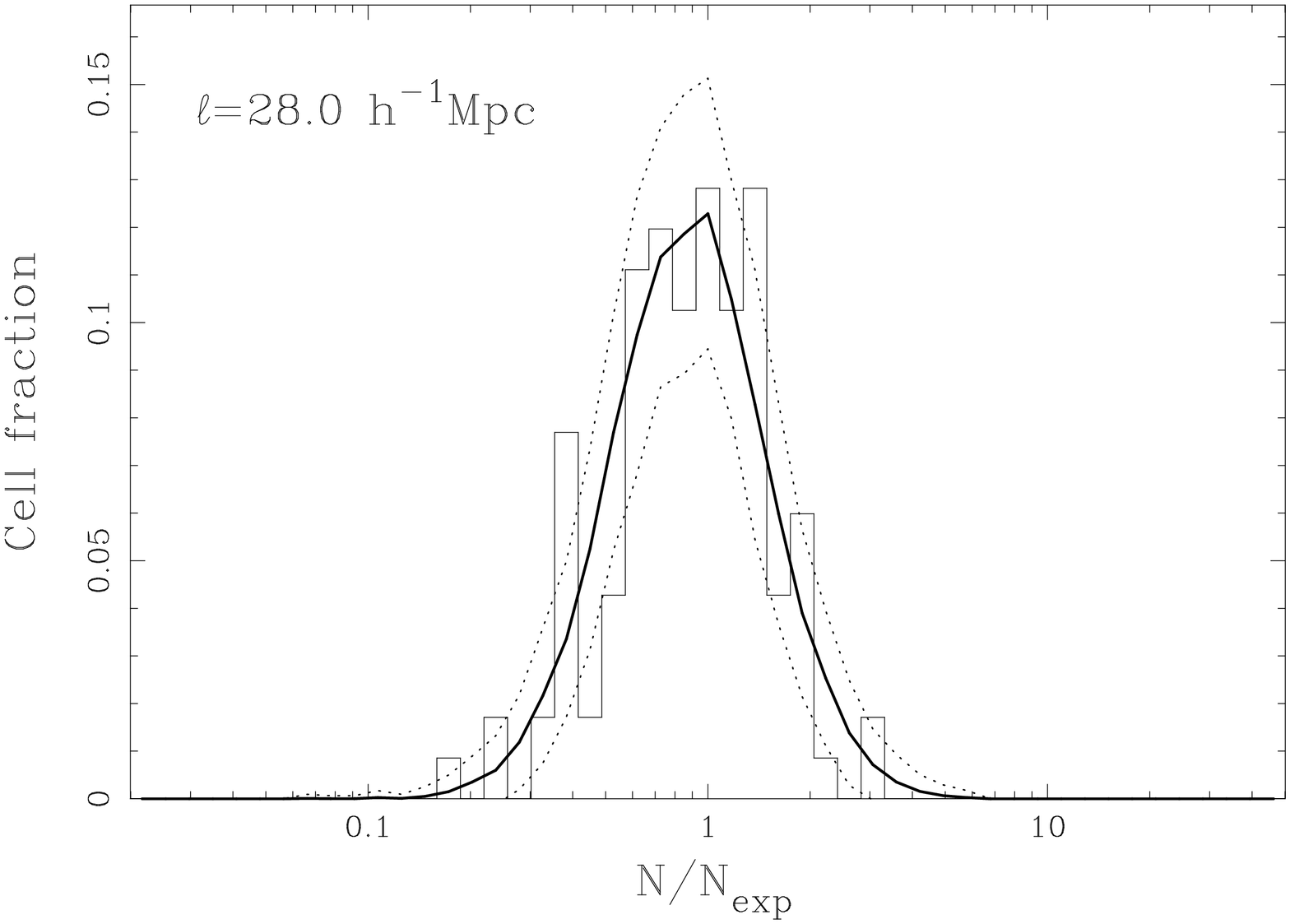}
\end{minipage}
\begin{minipage}[t]{0.33\textwidth}
\includegraphics[angle=-90, width=0.97\textwidth]{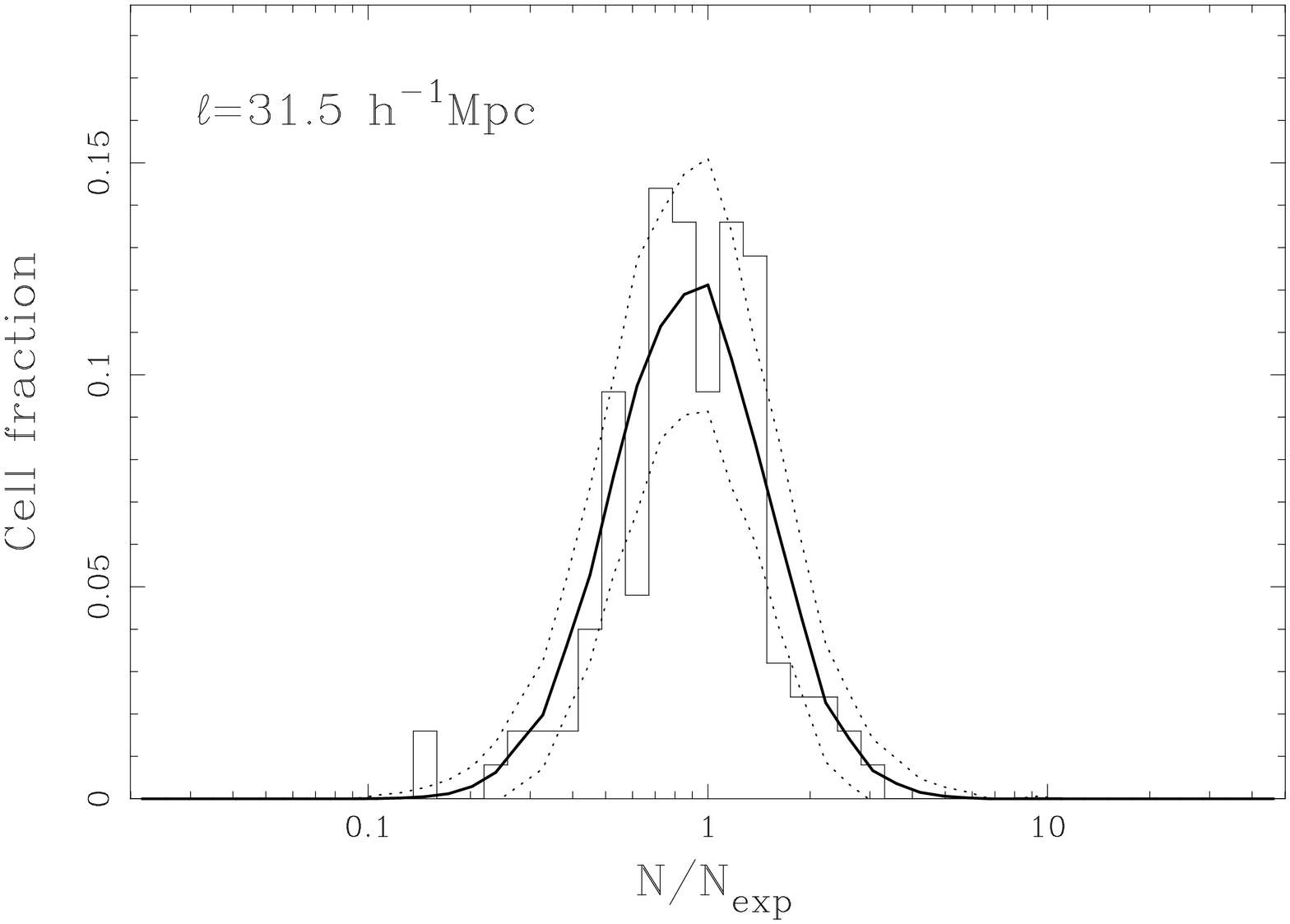}
\end{minipage}
\begin{minipage}[t]{0.33\textwidth}
\includegraphics[angle=-90, width=0.97\textwidth]{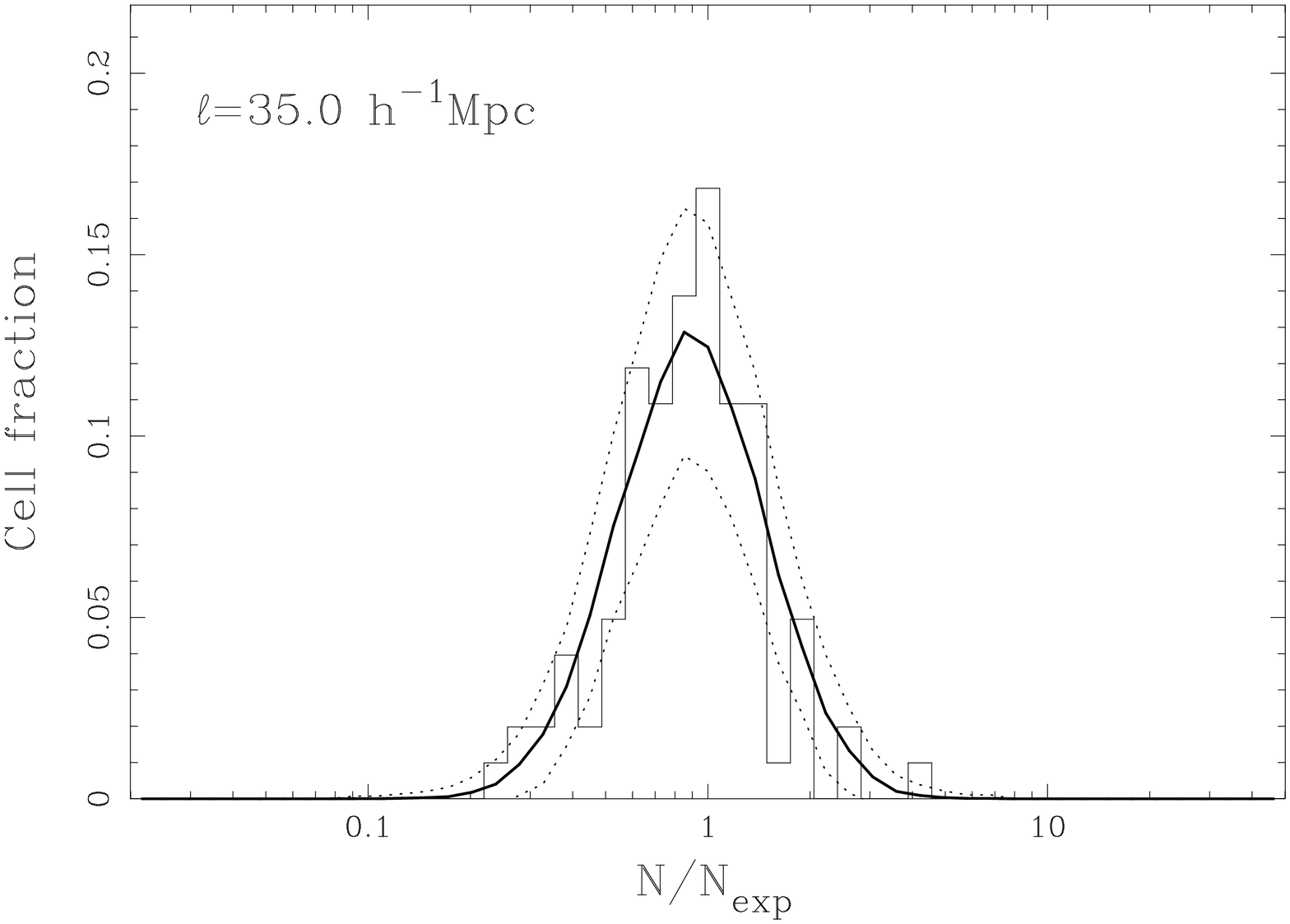}
\end{minipage}\hfill
\begin{minipage}[t]{0.33\textwidth}
\includegraphics[angle=-90, width=0.97\textwidth]{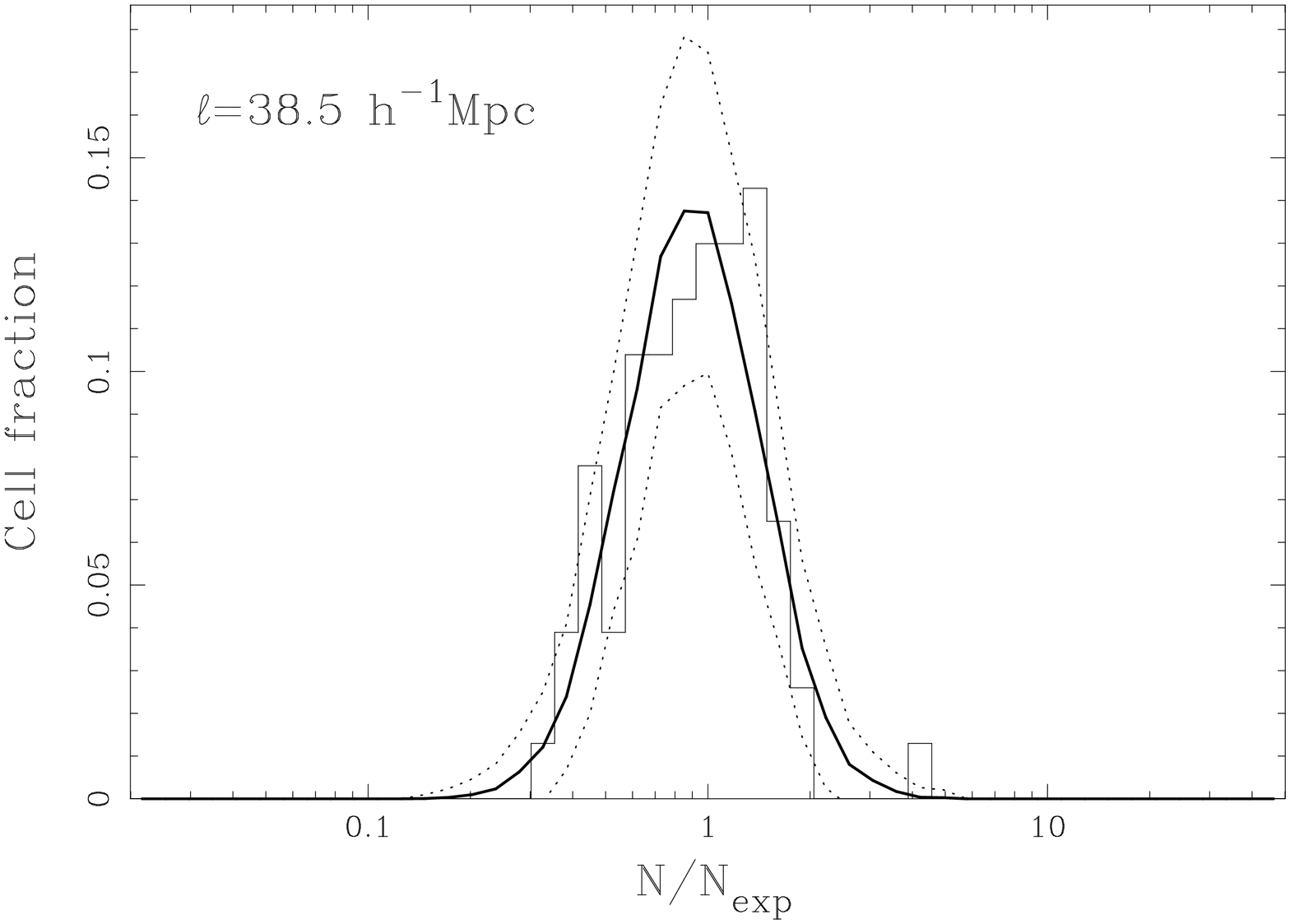}
\end{minipage}
\begin{minipage}[t]{0.33\textwidth}
\includegraphics[angle=-90, width=0.97\textwidth]{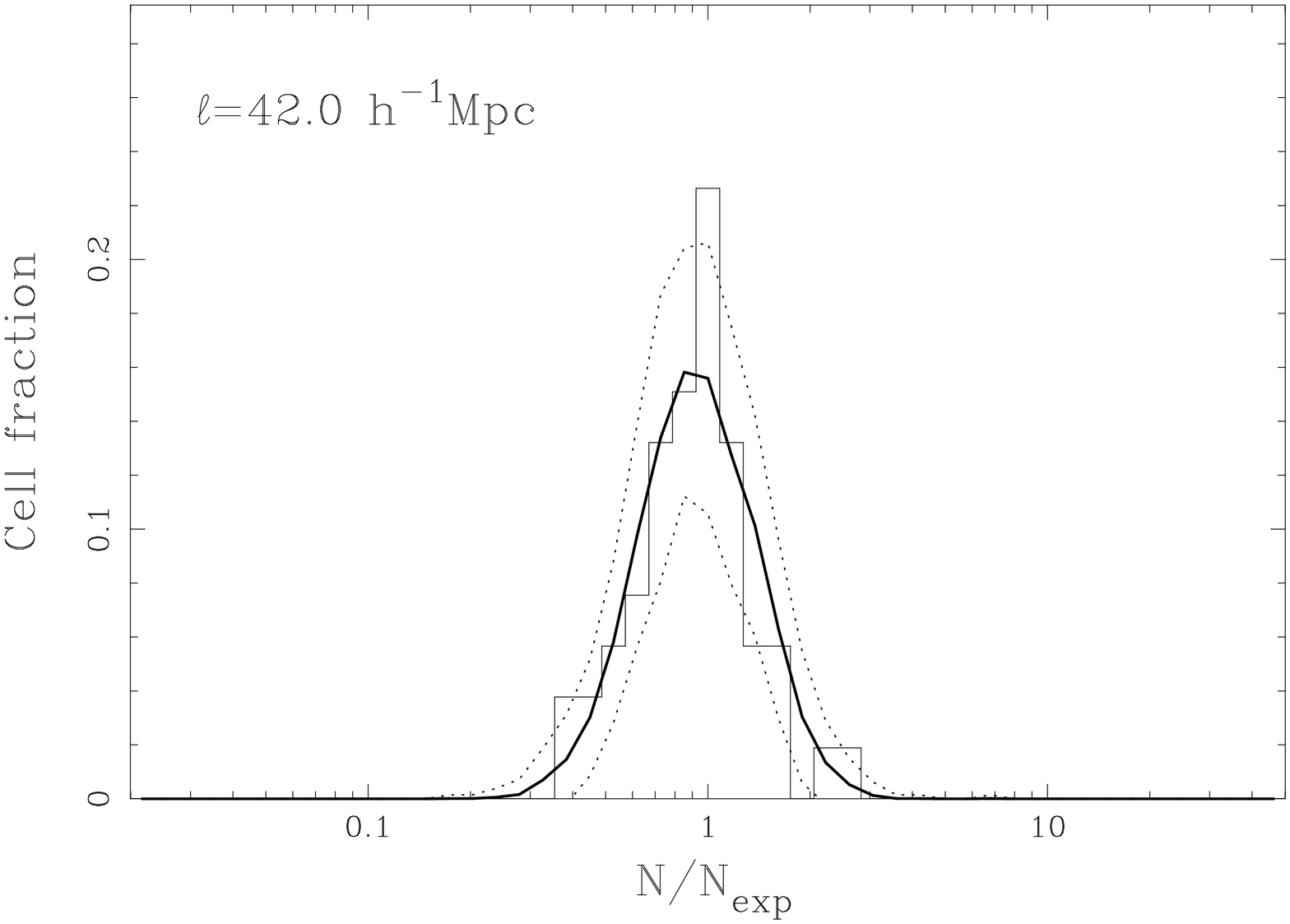}
\end{minipage}
\caption{The one-point distribution function for counts in cells of all galaxies with $\eta$ type for the SGP region over a range of cell sizes, from left--right and top--bottom, $\ell=14,17.5,21,24.5,28,31.5,35,38.5,42 h^{-1}$Mpc. The average of a large number of realizations of the best-fitting lognormal models, convolved with the same $N_{{\rm exp},i}$ as the data, are also shown, together with their 1-$\sigma$ spread. (Note that the $y$ axis changes between plots in this figure).}
\label{more_fN}
\end{minipage}
\end{figure*}

\begin{figure}
\begin{center}
\includegraphics[angle=-90, width=0.45\textwidth]{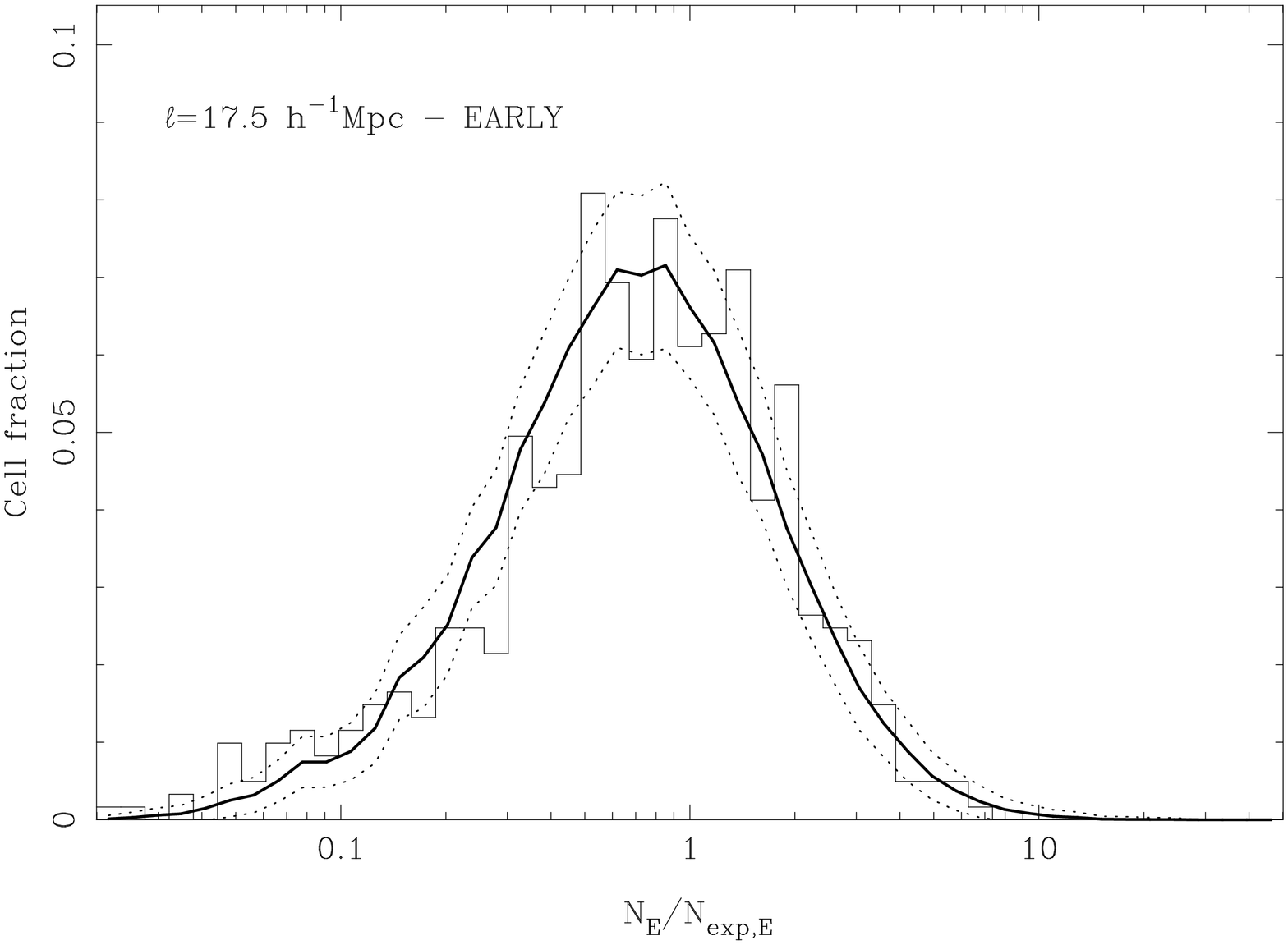}
\includegraphics[angle=-90, width=0.45\textwidth]{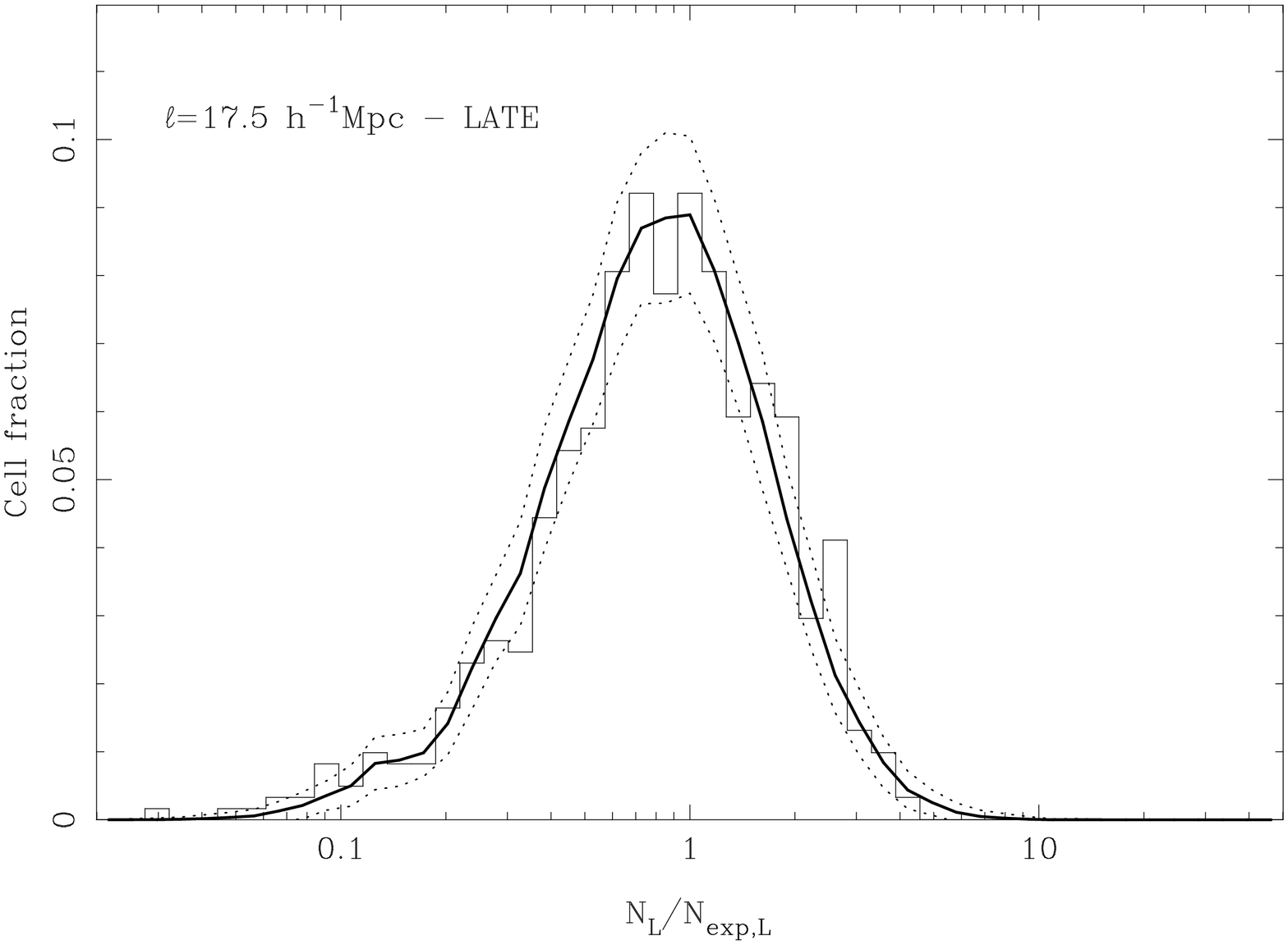}
\includegraphics[angle=-90, width=0.45\textwidth]{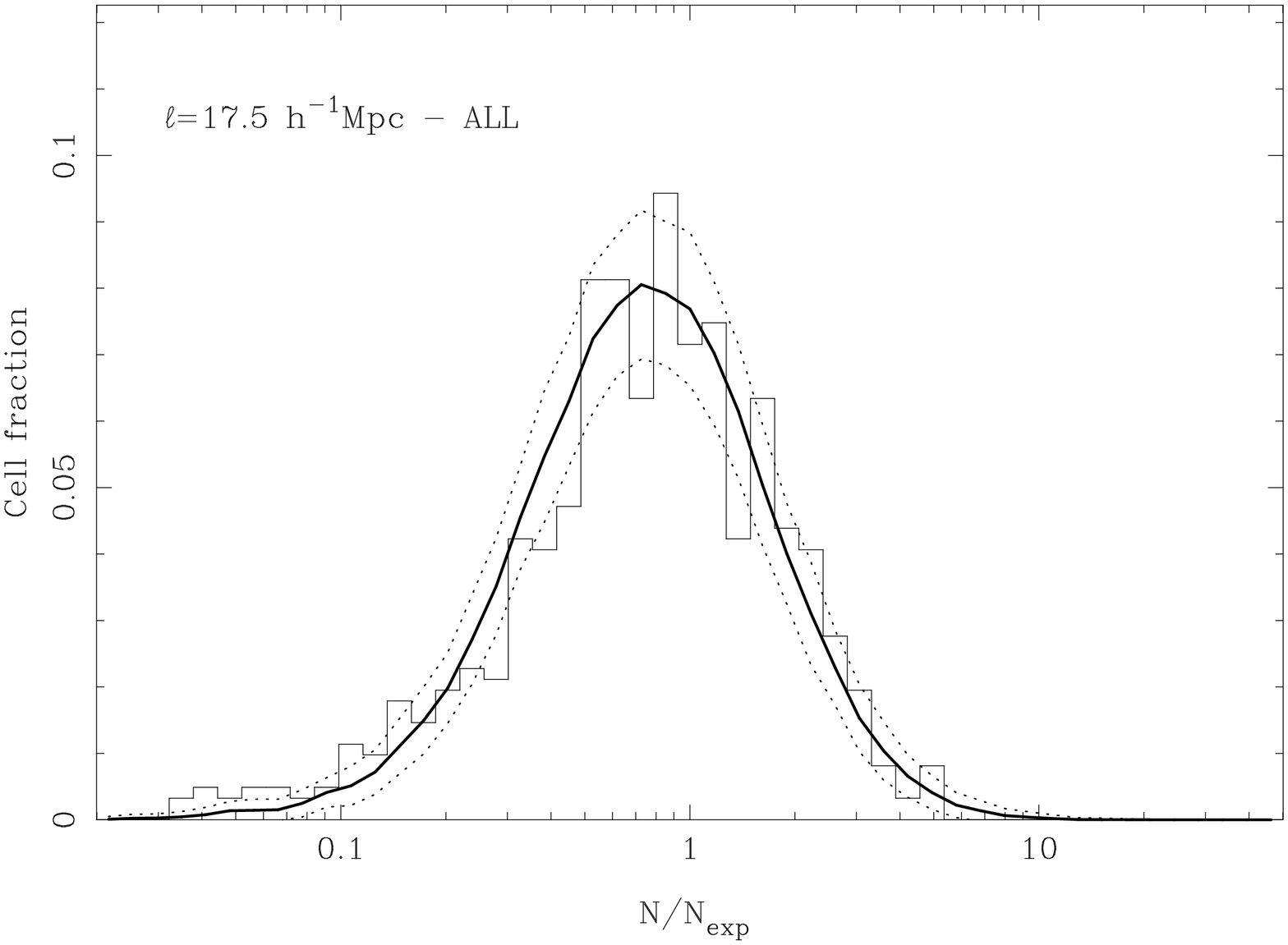}
\end{center}
\caption{An example of the one-point distribution function for counts
  in cells of size $\ell=17.5h^{-1}$Mpc in the NGP region for early-type
  galaxies (top), late types (middle), and all galaxies with $\eta$-types (bottom). The average of a large number of realizations of the best-fitting lognormal models, convolved with the same $N_{\rm exp,i}$ as the data, are also shown, together with their
  1-$\sigma$ spread. The higher variance of the early-type distribution is very clear.
}
\label{f_N_plots}
\end{figure}

\subsection{Lognormal model variances and empty cells}
\label{empty_cells}

\begin{table*}
\centering	
 \begin{minipage}{125mm}
  \caption{Number of empty cells in full survey data compared to the 10\% \& 90\% percentiles of the empty cells in lognormal models matching the measured variances in cells. The rightmost column shows the number of empty cells when the Hubble Volume mock catalogues are analysed in the same way.}
\begin{tabular}{lccccccc}
\hline
 & \multicolumn{2}{c}{early-type galaxies} & \multicolumn{2}{c}{late-type galaxies} & \multicolumn{2}{c}{all $\eta$-typed} & HV mocks \\
$\ell$ ($h^{-1}$Mpc) & $N_{\rm empty}$ & Models & $N_{\rm empty}$ & Models & $N_{\rm empty}$ & Models & $\left\langle N_{\rm empty} \right\rangle$ \\
\hline
7.0  & 14103  & 7482--7670   & 13495  & 8050--8237  & 10527  & 4243--4406 & 8029  \\
8.75 & 5238   & 2633--2743   & 4757   & 2803--2909  & 3405   & 1276--1355 & 2063 \\
10.5 & 1863   & 884--961     & 1590   & 962--1040   & 1005   & 354--405   & 478 \\
12.2 & 709    & 321--370     & 601    & 339--387    & 333    & 99--128    & 109  \\
14.0 & 211    & 89--114      & 163    & 88--113     & 83     & 19--31     & 17 \\

\hline
\label{Empties}
\end{tabular}
\end{minipage}
\end{table*}

We have fitted lognormal models for the one-point density distribution
to both the early- and late-type galaxy populations, and to the whole
sample. Examples of the actual one-point distributions for a cell size of
$\ell=17.5h^{-1}$Mpc for early and late types, as well as for the full data set (all galaxies with $\eta$ type) are shown in Fig.~\ref{f_N_plots}, and one-point distributions as a function of cell size $\ell$ for the SGP region are shown in Fig.~\ref{more_fN}. Fitting models to all cells including empty cells results in variances, particularly on small scales, which are many sigma above both the variance predicted from $\xi(s)$ and the measurements of the variance in cells of Section~\ref{variances}. For example, the best-fitting lognormal model for $\ell=10.5h^{-1}$Mpc gives a variance in early types of $\sigma_{E}\sim1.9$, whereas our previous measurements give $\sigma_{E}\sim1.1$. This discrepancy comes from the difficulty that the models have in reproducing the observed number of empty cells.

Some insight into why empty cells have such a dramatic effect can be gained by examining the number of empty cells in the data on scales where they become significant ($\ell\lesssim15h^{-1}$Mpc). Table~\ref{Empties} shows the number of empty cells in the full survey data for each type and for all $\eta$-typed galaxies, along with the 10\% and 90\% percentiles of the distribution of empty cells in a large number of Monte Carlo realizations of lognormal models with variances matching our previous measurements of the variance in cells. A Poisson-sampled lognormal model with a realistic variance cannot explain the number of empty cells in the data; the large excess of empty cells will increase the variance of the best-fitting lognormal model. Table~\ref{Empties} also shows the number of empty cells which are found when we analyse the hubble volume mock catalogues appropriately sampled to match the survey data (Cole \etal\ 1998; Norberg \etal\ 2002b) using the same method. The number of empty cells in the real and mock data on small scales exceeds the number predicted by the lognormal model, which suggests that the lognormal model is not a good fit to the actual density distribution function. In the real data this discrepancy is more pronounced and extends to smaller scales. This is related to the void probability function, as discussed in Croton \etal\ (2004a).

A simple solution to this problem is to fit lognormal models to the counts in cells excluding empty cells. Clearly this will cause variance estimates to be biased on scales where empty cells are significant ($\ell\lesssim15h^{-1}$Mpc). We have estimated the magnitude of this bias both by measuring the effect of excluding empty cells from Monte-Carlo realizations of lognormal models, and by considering the bias introduced into the variance measurements on small scales of Section~\ref{variances} when we excluded empty cells; both methods give similar results.

\begin{figure}
\begin{center}
\includegraphics[angle=-90, width=0.45\textwidth]{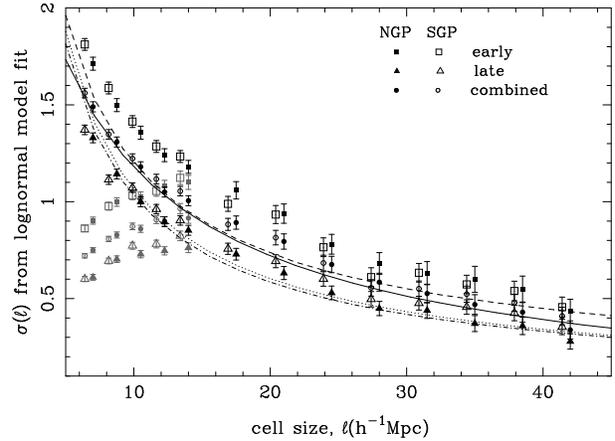}
\end{center}
\caption{$\sigma$ from the best fit lognormal model as a
  function of cell size, $\ell$. Filled symbols are for the NGP region, open symbols are SGP (offset as previously). The results shown are fits to the early-type
  galaxies (squares), late types (triangles), and to both types
  combined (circles). Predictions are overlaid as in Fig.~\protect\ref{efstat_sig_plot}. The results are based on Counts in Cells with empty cells removed from the analysis. The small scale results (for $\ell\leq14h^{-1}$Mpc) are corrected for the bias resulting from excluding empty cells which causes the variance to be under-estimated, as illustrated by the grey points (see text).
}
\label{sig_plot}
\end{figure}

The values of $\sigma$ from the lognormal fits to the early- and late-type subsets, as well as the full catalogue, are shown in Fig.~\ref{sig_plot}. The black points show the measurements excluding empty cells and corrected for the bias; points in grey for small $\ell$ show the original measurements illustrating the bias introduced by excluding empty cells in a similar manner to Fig.~\ref{efstat_sig_plot}. Since the likelihood function is well approximated by a $\chi^{2}$ distribution with one degree of freedom we have derived 1--$\sigma$ errors by considering the values for $\sigma_{\rm LN}$ at which $\mathcal{L} = \mathcal{L}_{\rm min} + 1$. We have also obtained Monte Carlo estimates of the errors by using the procedure outlined above to fit a large number of models generated by randomly drawing the galaxy density contrast, $\delta$, from a lognormal distribution with the best fit value of $\sigma$ and then generating model counts from a Poisson distribution with intensity $\lambda=(1+\delta)N_{\rm exp}$, using the same expected counts in cells as were calculated for the data. The magnitudes of errors using both methods are identical.

As another measure of the goodness of fit for these models we show in Table~\ref{KS_probs} probabilities obtained by the application of a Kolmogorov--Smirnov (KS) test to the distribution of $N/N_{\rm exp}$. The KS test is not ideal for a number of reasons; it is rather insensitive to variation in the tails of distributions, which is where we would expect the lognormal model will have the most difficulty in matching the data. Strictly speaking the estimate of $P_{\rm KS}$ which we use here is no longer valid once the data has been used to fix any free parameters of the model, although any effects should be small since the number of data points we use is very much larger than the number of free parameters.

The KS test probabilities indicate that the lognormal model is an acceptable fit to the data on large scales for both early and late types, as well as for all galaxies. None of the values for $P_{\rm KS}$ for $\ell\geq17.5h^{-1}$Mpc are sufficiently low to exclude the model at a high level of confidence. In general the early-type distribution is well fit by lognormal models to smaller scales than the distribution of late types or the combined galaxy distribution. On scales smaller than $\ell=10.5h^{-1}$Mpc a lognormal model is not a satisfactory fit to any of the distributions. We expect, based on the findings of Ueda \& Yokoyama (1996), that the lognormal model will not adequately describe the data in the non-linear regime. Our results are consistent with this expectation since the variance is significantly higher than unity on the scales at which the lognormal model becomes unsatisfactory.

\begin{figure}
\begin{center}
\includegraphics[angle=-90, width=0.45\textwidth]{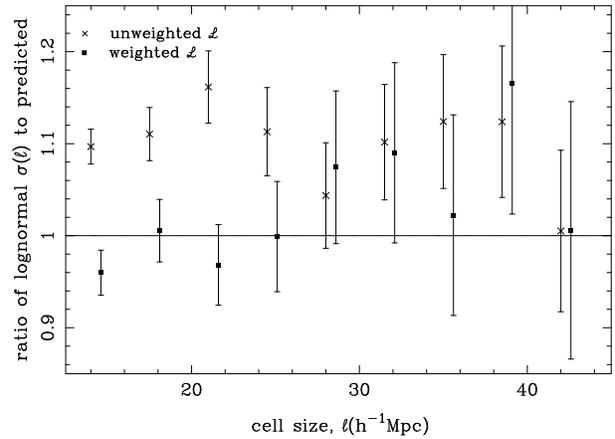}
\end{center}
\caption{Ratio plot of $\sigma(\ell)$ from the best fit lognormal model compared to the predicted value integrating $\xi(s)$ over cells. The crosses show the unweighted maximum likelihood results for all galaxies with $\eta$ types and NGP and SGP regions combined. Filled squares (offset) show the results when an additional weighting is applied to $\mathcal{L}(\sigma_{\rm LN})$ giving more weight to more overdense cells.
}
\label{ratio_plot}
\end{figure}

Although the values of $\sigma$ for all $\eta$ shown in Fig.~\ref{sig_plot} are broadly consistent with the cell variance derived from $\xi(s)$ in Section~\ref{predictions}, there is a systematic trend for the fitted values of $\sigma(\ell)$ to be higher than predicted. The magnitude of this effect is between 10--15\%, as shown in the ratio plot of Fig.~\ref{ratio_plot}; this corresponds to around a 2-$\sigma$ effect. Maximum likelihood fits to lognormal models giving variances consistent with predictions are obtained with the introduction of  an additional weighting factor to the likelihood defined in Eq.~\ref{curlyL}:
\begin{equation}
\mathcal{L}' = -2\sum_{i} \frac{N_{i}}{N_{{\rm exp},i}}\ln L_{i}.
\label{curlyL2}
\end{equation}

The results of applying the above weighting factor are shown by the filled squares in Fig.~\ref{ratio_plot}. In effect this modified likelihood gives more weight to the most dense regions, and suggests that the lognormal model is more appropriate to describe the density distribution of high density regions. This would be consistent with our observation that in general the early type distribution is better fit by a lognormal model than the distribution of late types, since early types are more prevalent in dense regions. We have tested the weighting scheme on model data based on a lognormal model and verified that it does not underestimate $\sigma$ in this case; the fact that a discrepancy such as this exists indicates that the lognormal model is not a completely satisfactory model for the one-point distribution functions. However, the KS test probabilities show that it is nevertheless an adequate prescription for our purposes, and indeed including the above weighting scheme does not alter our conclusions for the relative bias presented in the following section.

\section{The relative bias}\label{rel_bias}
We are now in a position to consider the characteristics of the joint distribution of the counts in cells. We postulate a smoothed density contrast field for galaxies $\delta_{g}$, which can be related to the density field of dark matter, $\delta$, using the general biasing framework of Dekel \& Lahav (1999)
\begin{equation}
\label{basic_bias}
\delta_{g}=b(\delta)\delta + \epsilon,
\end{equation}
which in principle is able to deal with both nonlinearity and stochasticity. We further assume that a similar relationship holds independently for the separate spectral types; in other words we consider early and late types with their own separate smoothed density fields denoted by $\delta_{E}$ and $\delta_{L}$ respectively. Then we can specify the relative bias between the density fields analogously to Eq.~\ref{basic_bias}:
\begin{equation}
\label{relative_bias}
\delta_{L}=b(\delta_{E})\delta_{E} + \epsilon.
\end{equation}

We have taken two approaches to quantifying the relative bias. Our first method considers an estimate of the galaxy density contrast for each spectral type in each cell, $i$, which we denote as
\begin{equation}
g_{E,i} = N_{E,i}/N_{E,{\rm exp},i},
\end{equation}
for the early-type galaxies, and analogously for late types.

Under the assumption that the density fields of both spectral types are related to the underlying dark matter field by a linear bias factor and that the only scatter is due to the Poissonian scatter caused by galaxy discreteness, we have for the early-type galaxies
\begin{equation}
\label{tb99_eq1}
g_{E,i}=b_{E}\delta_{i}+\epsilon_{E,i},
\end{equation}
where $\epsilon_{E,i}$ is the Poisson noise for the early types in
cell $i$. $g_{L,i}$ can be defined similarly. This is the basis for the null test described in Section~\ref{Teg99}.

In the second approach we attempt to fit to the joint distribution of the underlying smoothed density fields,
\begin{equation}
f(\delta_{E},\delta_{L}) = f(\delta_{L}|\delta_{E})f(\delta_{E}).
\label{blanton_eq2}
\end{equation}

We follow the example of Dekel \& Lahav (1999) and adopt a general description for relative bias:
\begin{equation}
b(\delta_{E})\delta_{E}\equiv\left\langle\delta_{L}|\delta_{E}\right\rangle=\int d\delta_{L}~f(\delta_{L}|\delta_{E})\delta_{L}.
\end{equation}
The function $b(\delta_{E})$ is characterized by Dekel \& Lahav (1999) by defining the following moments:
\begin{equation}
\label{b_moments}
\hat{b}\equiv\frac{\left\langle b(\delta_{E})\delta_{E}^{2}\right\rangle}{\sigma_{E}^{2}},\hspace{0.5cm}\tilde{b}^{2}\equiv\frac{\left\langle b^{2}(\delta_{E})\delta_{E}^{2}\right\rangle}{\sigma_{E}^{2}},
\end{equation}
where $\sigma_{E}\equiv\sqrt{\left\langle\delta^{2}_{E}\right\rangle}$, as we have used previously throughout this paper.

A random biasing field, $\bepsilon$, is defined in our case as 
\begin{equation}
\bepsilon\equiv\delta_{L}-\left\langle\delta_{L}|\delta_{E}\right\rangle,
\end{equation}
and the average biasing scatter, $\sigma_{b}$,
\begin{equation}
\label{sigma_b}
\sigma_{b}^{2}\equiv\frac{\left\langle\bepsilon^{2}\right\rangle}{\sigma_{E}^{2}}.
\end{equation}

These moments separate the effects of nonlinearity and stochasticity
of the bias relation. Linear bias is
often described by the ratio of variances of the density field. In
fact this bias parameter, $b_{\rm var}$, is a mixture of non-linear
and stochastic effects and can be expressed in terms of the above
moments as:
\begin{equation}
b_{\rm var}^{2}=\tilde{b}^{2}+\sigma_{b}^{2},\hspace{0.3cm}{\rm where}~ b_{\rm var}=\sigma_{L}/\sigma_{E}.
\end{equation}

The bivariate lognormal model considered by Wild et al (2004)
explicitly includes a stochastic term in the relative bias, which also
effectively introduces a non-linear term.

\subsection{Direct estimates of $b_{\rm var}$}

\begin{figure}
\begin{center}
\includegraphics[angle=-90, width=0.45\textwidth]{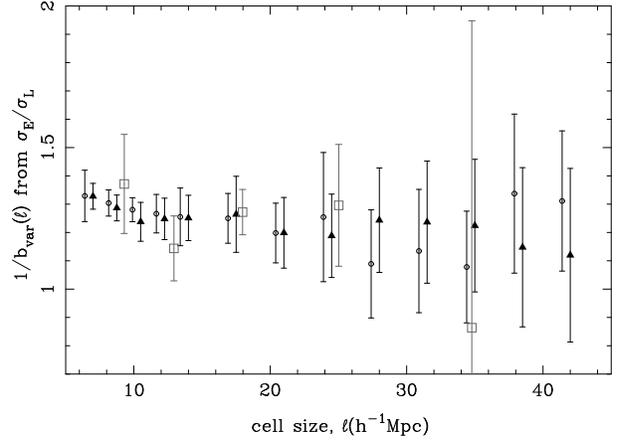}
\end{center}
\caption{The relative bias estimated from the ratio of $\sigma_{E}/\sigma_{L}$ from the variance estimator of Efstathiou et al.~(1990) for the NGP (filled triangles) and SGP (open circles) regions. The relative bias predictions from the real-space correlation functions per $\eta$ type of Madgwick et al.~(2003) are shown in grey.
}
\label{efstat_bias_plot}
\end{figure}

We estimated the relative bias from the calculated cell variances shown in Fig.~\ref{efstat_sig_plot_c}. We have used $1/b_{\rm var}$ to facilitate comparison with other papers where the relative bias is generally defined as a ratio $b_{\rm rel}=b_{E}/b_{L}$). The results are plotted in Fig.~\ref{efstat_bias_plot} which shows that the relative bias is consistent with a constant $1/b_{\rm var}=1.25\pm0.05$ for both the NGP and SGP regions and for all cell sizes. We have used only the $\ell>14h^{-1}$Mpc results in this estimate; for the error calculation we have used our measured error bars at each value of $\ell$ and made the assumption that adjacent bins in Fig.~\ref{efstat_sig_plot_c} are perfectly correlated. For comparison we have plotted the relative bias, $b_{\rm rel}=\sqrt{\xi_{E}(r)/\xi_{L}(r)}$, from the real-space correlation functions per $\eta$ type of Madgwick \etal~(2003b), where we have converted the separation, $r$, to $\ell$ by assuming $r\equiv R_{T}$ and using Eq.~\ref{scales}. This is intended mainly for illustration since, following the bias framework of Dekel \& Lahav (1999), the bias parameter formed from the ratio of correlation functions is not, in general, equivalent to $b_{\rm var}$. Clearly, though, they are consistent within the rather large errors from the correlation function estimate.

\begin{figure}
\begin{center}
\includegraphics[angle=-90, width=0.45\textwidth]{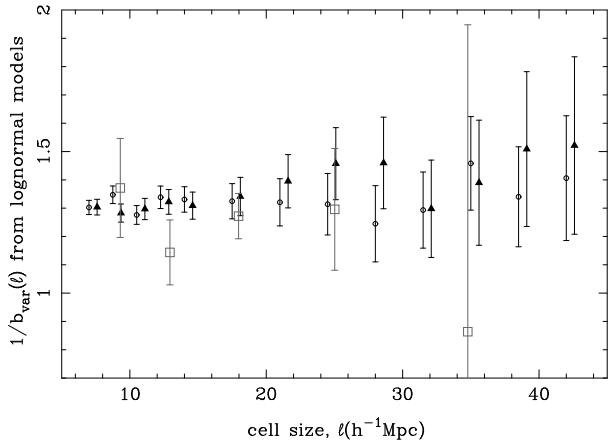}
\end{center}
\caption{The relative bias estimated from the parameters of the best fit lognormal models for the early- and late-type one-point distributions, for the NGP (filled triangles) and SGP (open circles) regions. The relative bias predictions from the real-space correlation functions per $\eta$-type of Madgwick et al.~(2003) are shown in grey.
}
\label{sig_bias}
\end{figure}

 We have also calculated the relative bias from the variances of the lognormal model fits to the early- and late-type one-point distributions as shown in Fig.~\ref{sig_plot},
\begin{equation}
b_{\rm var}=\sqrt{\frac{\exp(\sigma^{2}_{{\rm LN},E})-1}{\exp(\sigma^{2}_{{\rm LN},L})-1}}.
\end{equation}
The results are shown in Fig.~\protect\ref{sig_bias}, again compared to the results of Madgwick \etal~(2003b). The relative bias factor is again consistent with a scale invariant bias and we derive a value of $1/b_{\rm var}=1.28\pm0.05$ from the $\ell>14h^{-1}$Mpc data again assuming correlation of adjacent bins in Fig.~\ref{sig_bias}.

\subsection{The Tegmark `null-buster' test}\label{Teg99}

\begin{figure}
\begin{center}
\includegraphics[angle=-90, width=0.45\textwidth]{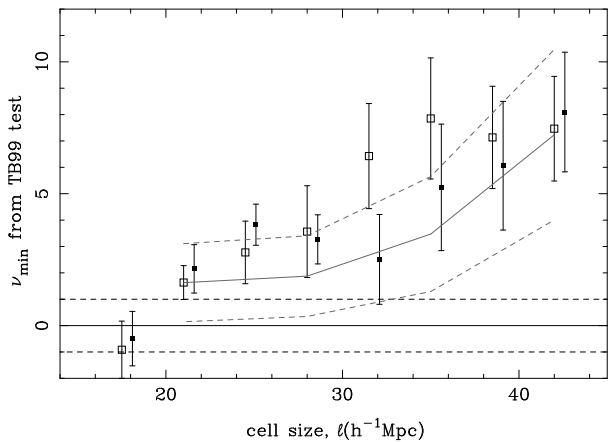}
\end{center}
\caption{Variation of ($\nu_{\rm min}$) from the TB99 test
  with scale. The solid and dashed lines show the expected value and
  1-$\sigma$ variation for results consistent with a linear bias relation. The solid and open squares show the average $\nu_{\rm min}$ and its 1-$\sigma$ scatter measured over a number of separate cell divisions obtained by shifting the original divisions, for the NGP and SGP respectively. The grey solid and dashed curves show respectively the average and 1-$\sigma$ spread of $\nu_{\rm min}$ for models including the effects of the selection function variations on scales $<\ell$.
}
\label{nu_min}
\end{figure}

\begin{figure}
\begin{center}
\includegraphics[angle=-90, width=0.45\textwidth]{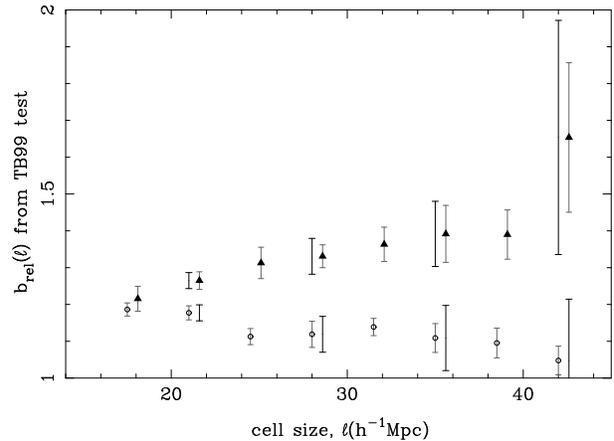}
\end{center}
\caption{The value of the relative bias factor, $b_{\rm rel}$, giving the minimum value for $\nu$ in the TB99 test. The solid triangles and open circles show the average $b_{\rm rel}$ and its 1-$\sigma$ scatter measured over a number of separate cell divisions obtained by shifting the original divisions, for the NGP and SGP respectively. The black error bars are a more realistic estimate for the errors based on models including the effects of selection function variations on scales $<\ell$.
}
\label{tb99_bias}
\end{figure}

\citet{Tegmark_1999} describes a simple `null buster' test, based on a generalized $\chi^{2}$ statistic, to rule out the possibility that the density fields traced by galaxies of two different spectral types can be related by a simple deterministic linear biasing prescription. This test has been used by Seaborne \etal\ (1999) to compare the PSCz and Stromlo-APM redshift surveys.

If we first assume that the estimated galaxy density contrasts for each spectral type, $g_{E}$ and $g_{L}$, are related to an underlying dark matter density field by the prescription of Eq.~\ref{tb99_eq1}, then we can construct the difference map:
\begin{equation}
{\bf \Delta g} \equiv {\bf g}_{E} - f{\bf g}_{L},
\end{equation}
for different values of the relative bias factor $f = b_{E}/b_{L}$.

If the deterministic linear bias model is valid, then for the correct value of $f$, the relative bias factor, ${\bf \Delta g}$
will consist merely of Poisson noise, which will have a covariance
matrix given by \citet{Tegmark_1999b} as
\begin{equation}
\label{noise}
{\bf N} \equiv \left\langle {\bf \Delta g \Delta g}^{t} \right\rangle = \delta_{ij}\left[ 1/N_{E,{\rm exp},i} + f^{2}(1/N_{L,{\rm exp},i}) \right].
\end{equation}

Since we are testing the null hypothesis that $\left\langle {\bf
  \Delta g \Delta g}^{t} \right\rangle = {\bf N}$, we can define
  $\chi^{2}={\bf \Delta g}^{t}{\bf N}^{-1}{\bf \Delta g}$. If the null
  hypothesis is correct, the quantity
  $\nu=(\chi^{2}-N_{c})/\sqrt{2N_{c}}$, where $N_{c}$ is the number of
  cells, has an expectation of zero and standard deviation of one. We
  can therefore interpret $\nu$ as a measure of the significance with
  which the null hypothesis is ruled out.

In the case where there is extra signal, ${\bf S}$, in the covariance matrix of the difference map, so that $\left\langle {\bf \Delta g \Delta g}^{t} \right\rangle = {\bf N}
+ {\bf S}$, the generalized $\chi^{2}$ statistic \citep{Tegmark_1999} is a more powerful way to rule out the null hypothesis:

\begin{equation}
\label{mod_chi}
 \nu \equiv \frac{\mathbf{ \Delta g^{t}N}^{-1}\mathbf{ SN}^{-1}\mathbf{ \Delta g} -\;\mathrm{ Tr}\; (\mathbf{ N}^{-1}\mathbf{ S})}{[2\;\mathrm{ Tr}\;(\mathbf{ N}^{-1}\mathbf{ SN}^{-1}\mathbf{ S})]^{1/2}}.
\end{equation}

If there are any deviations from deterministic linear bias we would
expect these to be correlated with large scale structure. We therefore
choose the matrix ${\bf S}$ to be the covariance between cell
overdensities calculated using the redshift-space correlation function
calculated by \citet{Hawkins_2003}, i.e.~the volume average of
$\xi(s_{ij})$ over cells $i$ and $j$. The value of $\nu$ depends only
on the shape of $\mathbf{S}$, not on its amplitude.

Note that these tests are only valid when fluctuations are close to
Gaussian. For this reason, when we apply this test we exclude cells
where $g_{E,i} > 1$ or $g_{L,i} > 1$. The test also uses a Gaussian
approximation for the Poisson fluctuations described by the covariance
matrix ${\bf N}$, so we apply an additional cut on cells where this
will be particularly inaccurate, where $N_{\rm exp}\leqslant 10$. This
value for the cut in $N_{\rm exp}$ is a compromise between a value
which renders the error term from using the Gaussian approximation
negligible and the necessity of not removing too many cells. Even so
we find that the test is not applicable for
$\ell\leq17.5h^{-1}$Mpc. We have reduced the redshift range to $0.03<z\leq0.12$ for the purposes of this test in order
that our cut in $N_{\rm exp}$ does not introduce systematic effects.

% Another effect that may contribute to the apparent signal is 
% the uncertainty in the mask. The noise covariance matrix
% (Eq.~\ref{noise}) depends on knowing $N_{\rm exp}$ for each
% cell, so clearly an error in the assumed $N_{\rm exp}$ can
% yield an incorrect noise estimate. 
% A cell that is more incomplete than we assume must contain a larger
% true number of total galaxies, so the Poisson fluctuations in the
% early:late ratio will be smaller than we assume.
% Conversely a cell that is more complete than we assume must contain a
% smaller true number of total galaxies, so the Poisson fluctuations in the
% early:late ratio will be larger than we assume.
% When averaged over all cells the mean fluctuations will be
% approximately unbiassed, and so this is likely to be a small effect.

The minimum values of $\nu$ for a range of values for the cell size
$\ell$ are shown in Fig.~ \ref{nu_min}. We also plot error bars which
are derived from applying the test to cell divisions which are shifted
by up to $\ell/\sqrt{2}$ parallel to the ${\rm ra}$--$z$ plane, relative
to the original division. Clearly the error bars resulting from such
an approach will underestimate the true errors. In general the values
of $\nu_{\rm min}$ are not consistent with a linear and deterministic
relative bias; indeed the significance of this detection increases as
$\ell$ increases.

The apparent detection of stochasticity at such large scales should be
treated with some caution. The large cells contain a large number of
galaxies so the shot noise is small, but the results will be very
sensitive to any subtlety in the model and any small systematic error
in the data.  A simple possibility may be that the Poisson sampling
hypothesis of the model may not be exactly correct. An example of an
instrumental effect that could lead to an apparent detection of
stochasticity on large scales is an interaction between small-scale
stochasticity, and the variation of the selection function on scales
smaller than the cell division. This can produce an excess variance
over the covariance matrix assumed on the Tegmark test
(Eq.~\ref{noise}).  If the null hypothesis of the Tegmark test were
true on all scales we would expect $\nu_{\rm min}$ to be consistent
with zero.  However, there are differences in the distribution of
early and late-type galaxies on small scales, as seen in the
morphology-density relation.  Also there are small-scale variations in
both in the angular variations quantified by the survey mask and in
the radial variations in the $n(z)$.  So, it is possible that galaxies
of one type may preferentially reside in a region in a cell where the
selection function differs significantly from the cell average. After
allowing for incompleteness, the observed covariance of
${\mathbf{\Delta g}}$ will be enhanced relative to what would be
expected from simple Poisson noise as described by Eq.~\ref{noise}.

We have considered a simple model which includes small-scale
stochasticity, and variations in the selection function on scales less
than $\ell$, and found that it can reproduce our results from the
Tegmark test.  We first generate linear bias models matching the data
by drawing $\delta_{E}$ from the best-fitting lognormal model and
applying a linear bias function to obtain $\delta_{L}$. We then
generate a parent number of galaxies in each cell by Poisson sampling
the density field with a constant sampling rate assuming all the cells
are 100\% complete and with $n(z)$ set to be a constant equal to the
maximum $n(z)$ at the mean redshift of the survey. The parent galaxies
are distributed within each cell using a modified Rayleigh-L\'evy
flight model (see Peebles 1980, section 62) matching the correlation
function. We have modified the original Rayleigh-L\'evy flight model
so that the lacuniarity of the process is more realistic -- in effect
the voids in our clustered point process are less empty. We then
select or reject the parent galaxies based on the selection function
at the location of each galaxy. The point processes for model early
and late types are independent.

The grey solid and dashed lines in Fig.~ \ref{nu_min} show the average
and 1-$\sigma$ spread of $\nu_{\rm min}$ when the Tegmark test is
applied to our Rayleigh-L\'evy flight models. It can be seen that
including the effect of selection function variation on scales less
than the cell size can reproduce the kind of results seen in the data,
without the need to invoke any non-linear or stochastic relative bias
on large scales.

Nevertheless, a significant result from the Tegmark test means that it
is difficult to avoid the conclusion that nonlinearity and/or
stochasticity exists in the density field at some scale. We have
however shown that the detection of this effect on a given cell scale
can suffer from `aliasing' of the effect from sub-cell scales. Our
model for this effect assumes a relatively large stochasticity on
small scales, which may be unrealistic at least for large $\ell$
cells. It is an open question whether a more realistic model, for
example one based on the observed morphology-density relation, would
give the same effect; this is however beyond the scope of the current
paper.

We plot $f$ from the best-fitting linear bias models in
Fig.~\ref{tb99_bias}, again with errors derived from cell shifts. Note
that the values of $f$ from this test are not strictly comparable to
the other values quoted in this paper except in the case where a
deterministic linear bias model is an exact representation of the
data, since we have of necessity imposed a cut on
$\delta_{E}$. However we expect that on large scales this
approximation will be close enough for comparison to be
instructive. If we use the slightly larger error bars derived from our
Monte-Carlo realizations of Rayleigh-L\'evy flight models, and assume
measurements of $f$ in adjacent bins are correlated, we obtain
$f=1.28\pm0.03$ for the NGP and $f=1.16\pm0.03$ for the SGP. It is
notable that the best fit linear bias factors for the NGP and SGP
regions do not seem to be consistent. Averaging the two regions we
find a value for $f$ which is consistent with our measurements of
$b_{\rm var}$ presented in the previous section. A more relevant
comparison is with the maximum likelihood measurements of the linear
bias parameter, which we present in the following section, where we find
a similar discrepancy between NGP and SGP regions which we discuss
more fully in Section~\ref{last_bit}.

\subsection{Fitting the joint counts in cells}\label{jointCiC}
On the relatively large scales studied here, the results of the
Tegmark test can be made consistent with the density fields of early- and
late-type galaxies being related by a deterministic linear bias, once
variations in the selection function on scales smaller than the cell
size are considered. However, the test does not reveal any further
details of the nature of the relative bias between galaxies of
different spectral type. Blanton (2000) describes a more direct approach to measuring the relative bias and applies it to the Las Campanas Redshift Survey
(LCRS). The basis of this approach is a maximum-likelihood fit to the
joint counts in cells, $P(N_{E},N_{L})$, which is simply the joint
probability of the density fields convolved with Poisson
distributions.

If we convolve Eq.~\ref{blanton_eq2} with the expected Poissonian scatter we derive the following joint probability for the counts:
\begin{eqnarray}
P(N_{E},N_{L}) & = & \int d\delta_{E}~\frac{\lambda_{E}^{N_{E}}}{N_{E}!}e^{-\lambda_{E}}f(\delta_{E}) \nonumber\\
&\times & \int d\delta_{L}~\frac{\lambda_{L}^{N_{L}}}{N_{L}!}e^{-\lambda_{L}}f(\delta_{L}|\delta_{E}),
\label{blanton_eq3}
\end{eqnarray}
where 
\[
\lambda_{E}=N_{E,{\rm exp}}(1+\delta_{E}),
\]
and similarly for the late types.
% \begin{eqnarray}
% P &\hspace{-0.3cm} = &\hspace{-0.4cm} \int
% d\delta_{E}~\frac{N^{N_{E}}_{E,{\rm exp}}(1+\delta_{E})^{N_{E}}}{N_{E}!}e^{-N_{E,{\rm exp}}(1+\delta_{E})}f(\delta_{E})\nonumber\\ 
% && \hspace{-30pt}\times \int
% d\delta_{L}~\frac{N^{N_{L}}_{L,{\rm exp}}(1+\delta_{L})^{N_{L}}}{N_{L}!}e^{-N_{L,{\rm exp}}(1+\delta_{L})}f(\delta_{L}|\delta_{E}).\
% \label{blanton_eq3}
% \end{eqnarray}

We have used Eq.~\ref{blanton_eq3} to define the likelihood as a function of $\sigma_{{\rm LN},E}$ and a model for the relative bias $f(\delta_{L}|\delta_{E})$. We then found the maximum likelihood using a downhill simplex method (Press et al.~1992) and hence estimate the best-fitting relative bias.

 \subsubsection{Deterministic bias models}\label{bias}

\begin{figure}
\begin{center}
\includegraphics[angle=-90, width=0.45\textwidth]{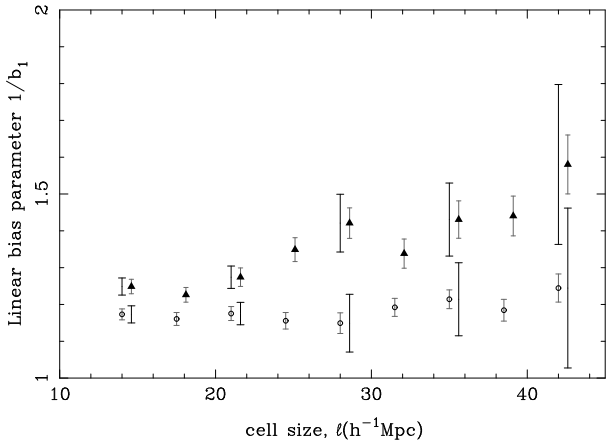}
\end{center}
\caption{The best-fitting linear bias parameter, $b_{E}/b_{L}=1/b_{1}$, as
  a function of cell size $\ell$ for the NGP (filled triangles) and SGP (open circles) regions. The grey error bars are derived by considering the value of $b_{1}$ for
  which $\mathcal{L} = \mathcal{L}_{\rm min} + 1$. Black error bars adjacent to selected points are a more realistic error estimate showing the effect on the errors of variation of the selection function on scales less than $\ell$.
}
\label{lin_plot}
\end{figure}

\begin{figure}
\begin{center}
\includegraphics[angle=-90, width=0.45\textwidth]{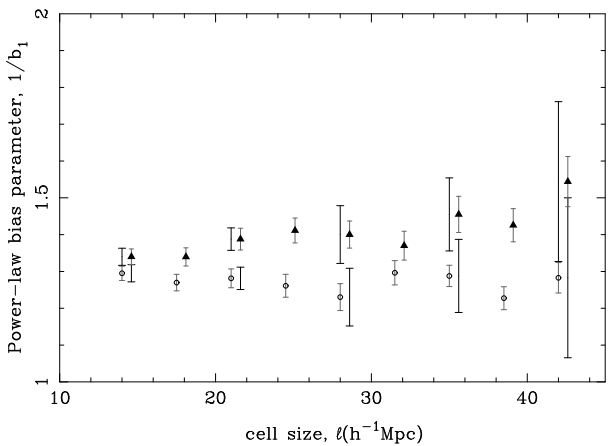}
\end{center}
\caption{The best-fitting power-law bias parameter, $1/b_{1}$, as
  a function of cell size $\ell$ for the NGP (filled triangles) and SGP (open circles) regions. The grey error bars are derived by considering the value of $b_{1}$ for
  which $\mathcal{L} = \mathcal{L}_{\rm min} + 1$. Black error bars adjacent to selected points are a more realistic error estimate showing the effect on the errors of variation of the selection function on scales less than $\ell$.
}
\label{pl_plot}
\end{figure}

The conditional density distribution function,
$f(\delta_{L}|\delta_{E})$, can in principle describe completely the
relationship between the galaxy density fields, including any
nonlinearity or stochasticity. In this paper we have concentrated on bias
models in which the density field of the late types is related to that of
the early types in a deterministic manner, i.e.~$f(\delta_{L}|\delta_{E})$ can be expressed in the form
\begin{equation}
f(\delta_{L}|\delta_{E})=\delta_{\mathrm D}[\delta_{L} -
b(\delta_{E})\delta_{E}],
\end{equation}
where $\delta_{\mathrm D}$ is the Dirac delta function. The possibility that the bias relation may exhibit additional scatter above the Poisson fluctuations is considered by Wild \etal\ (in preparation).

The simplest form for the bias relation is 
\begin{equation}
b(\delta_{E})\delta_{E}=b_{0} + b_{1}\delta_{E},
\label{lin_bias}
\end{equation}
corresponding to linear bias. Of course, this model gives
unphysical values for $\delta_{L}$ when $b_{1}>1$ and
$-1\leqslant\delta_{E}<0$; in this case we set $b(\delta_{E})=0$, following the example of Blanton (2000).

A simple generalization of the bias relation which includes non-linear
effects is a power-law bias model,
\begin{equation}
b(\delta_{E})\delta_{E}=b_{0}(1+\delta_{E})^{b_{1}} - 1.
\label{Pl_bias}
\end{equation}
For both these models there is only one free parameter ($b_{1}$),
since $b_{0}$ is set by the requirement $\left\langle \delta_{L}
\right\rangle = 0$.

% At the cost of adding an additional free parameter, we have also
% considered a modified power-law bias model, which includes a linear
% term:
% \begin{equation}
% b(\delta_{e})=b_{0}(1+b_{2}\delta_{e})^{b_{1}} - 1.
% \label{ModPl_bias}
% \end{equation}

We have fitted deterministic bias models to the counts in cells only
for $\ell\geq14h^{-1}$Mpc since below this scale the problem of empty
cells becomes significant. The parameters for the best fit linear bias
model over a range of cell sizes are shown in Fig.~ \ref{lin_plot}
(where we have used $1/b_{1}$ since this corresponds to what is
normally understood as the relative bias, namely $b_{\rm
rel}=b_{E}/b_{L}$, where $b_{E},b_{L}$ correspond to the linear bias
factors for early and late types). 

There is a systematic trend for the NGP early types to be more
strongly biased than in the SGP. Such a discrepancy, also seen in the
results from the modified $\chi^{2}$ test, was not observed in the
variance measurements so it is important to consider whether the
effect is indeed as significant as it would appear. The error bars
given in these plots are from the likelihood as a function of $b_{1}$
for the model, and do not include potential correlation between $b_{1}$
and $\sigma_{E}$ which could lead to them being
underestimated. Examination of the two dimensional likelihood contours
reveals that in fact the bias parameter and $\sigma_{E}$ are not
significantly correlated, so neglecting $\sigma_{E}$ will not lead to
an underestimation of the errors in $b_{1}$. We have also calculated
errors by fitting Monte Carlo realizations of the bias models
including our measured errors in $\sigma_{E}$. These are identical in
magnitude to the errors obtained from the likelihood function, again
suggesting that any correlation between $b_{1}$ and $\sigma_{E}$ is
not significant.

If we repeat our Monte Carlo error analysis using the Rayleigh-L\'evy
flight models with a selection function that varies on scales less
than $\ell$, as in Section~\ref{Teg99}, the actual errors become rather
larger. If we assume that measurements of $b_{1}$ in adjacent bins of $\ell$ are correlated we obtain $b_{1,{\rm lin}}=1.27\pm0.04$ for the NGP and $b_{1,{\rm lin}}=1.17\pm0.04$ for the SGP, which corresponds to just less than a 2-$\sigma$ discrepancy. 

The variation of the best-fitting power-law bias parameter (again
using $1/b_{1}$ for consistency) with cell size $\ell$ is shown in
Fig.~\ref{pl_plot}. If we again assume that measurements of $b_{1}$ in
adjacent bins of $\ell$ are correlated, we obtain $b_{1,{\rm
PL}}=1.36\pm0.05$ for the NGP and $b_{1,{\rm PL}}=1.29\pm0.04$ for the
SGP. These results are noticeably higher than the linear bias
parameters, showing that the assumption of linear bias pushes estimates
of the bias parameter closer to unity to compensate for nonlinearities
in the data. Once we account properly for non-linear biasing the bias
parameters approach consistency between regions.

\begin{figure*}
\begin{minipage}[t]{0.5\textwidth}
\includegraphics[angle=-90, width=0.97\textwidth]{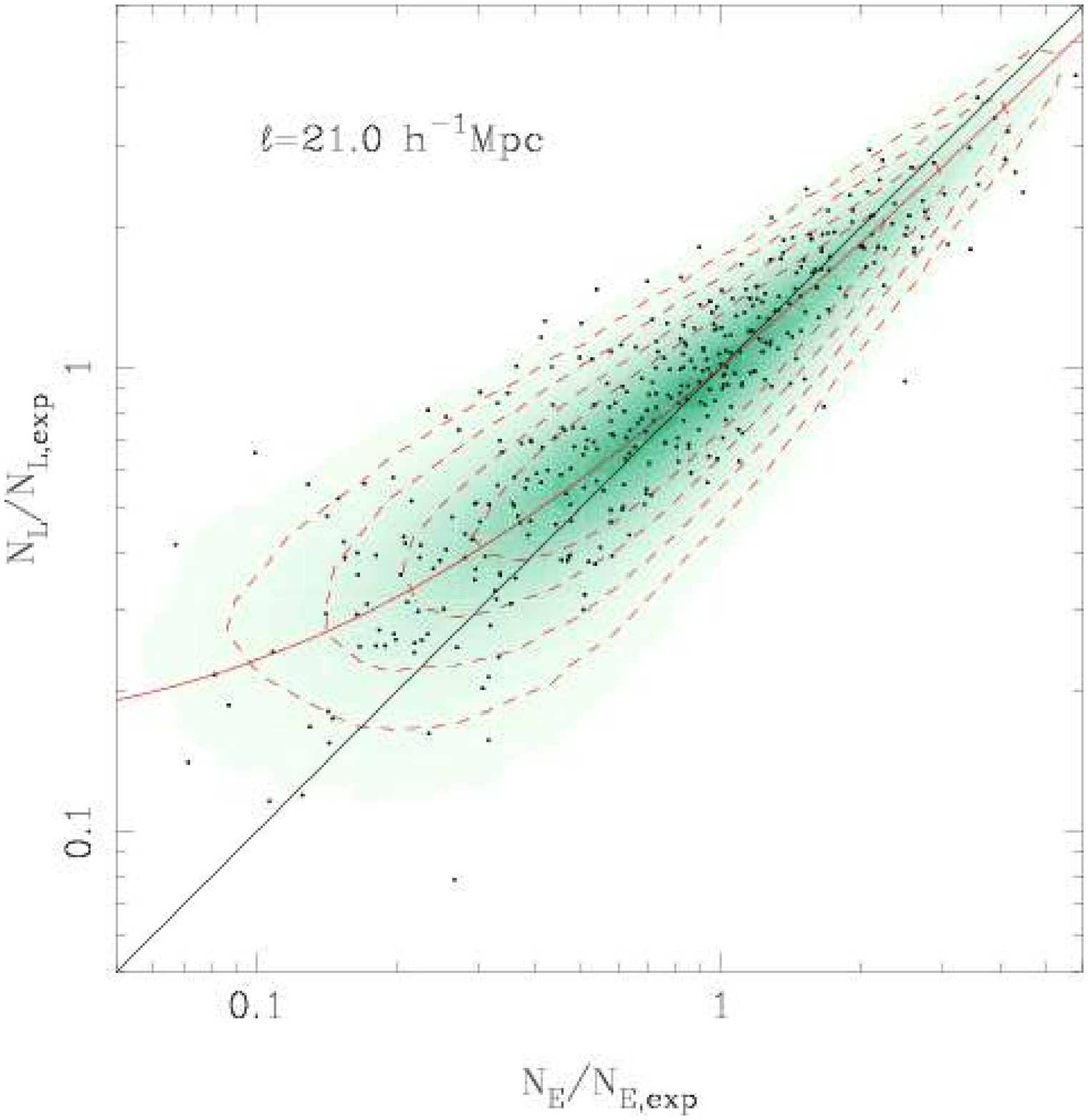}
\end{minipage}\hfill
\begin{minipage}[t]{0.5\textwidth}
\includegraphics[angle=-90, width=0.97\textwidth]{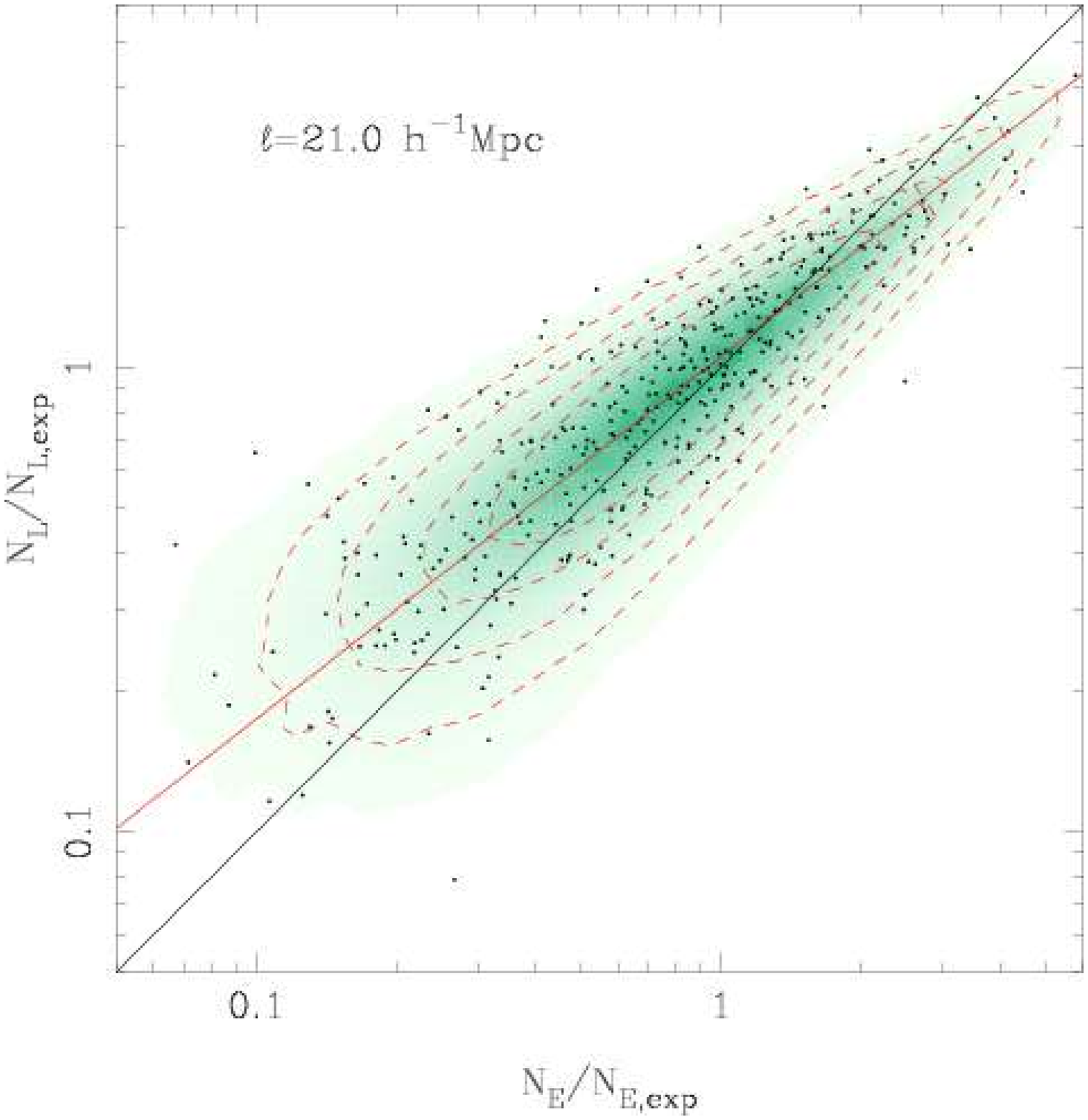}
\end{minipage}
\caption{An example of the joint counts in cells for the SGP region with $\ell=21h^{-1}$Mpc. The colour scale and contour levels are derived from Monte Carlo realizations of the best fit linear bias model (left) and the best fit power-law bias model (right), using the same expected counts as the data, and the dashed lines indicate the 50\%,70\%,85\% and 93\% significance levels. The black solid line indicates a mean relative bias of 1 and the red line shows the mean relative bias for the model.
}
\label{cont_plot}
\end{figure*}

\begin{figure*}
\begin{minipage}[t]{0.33\textwidth}
\includegraphics[angle=-90, width=0.97\textwidth]{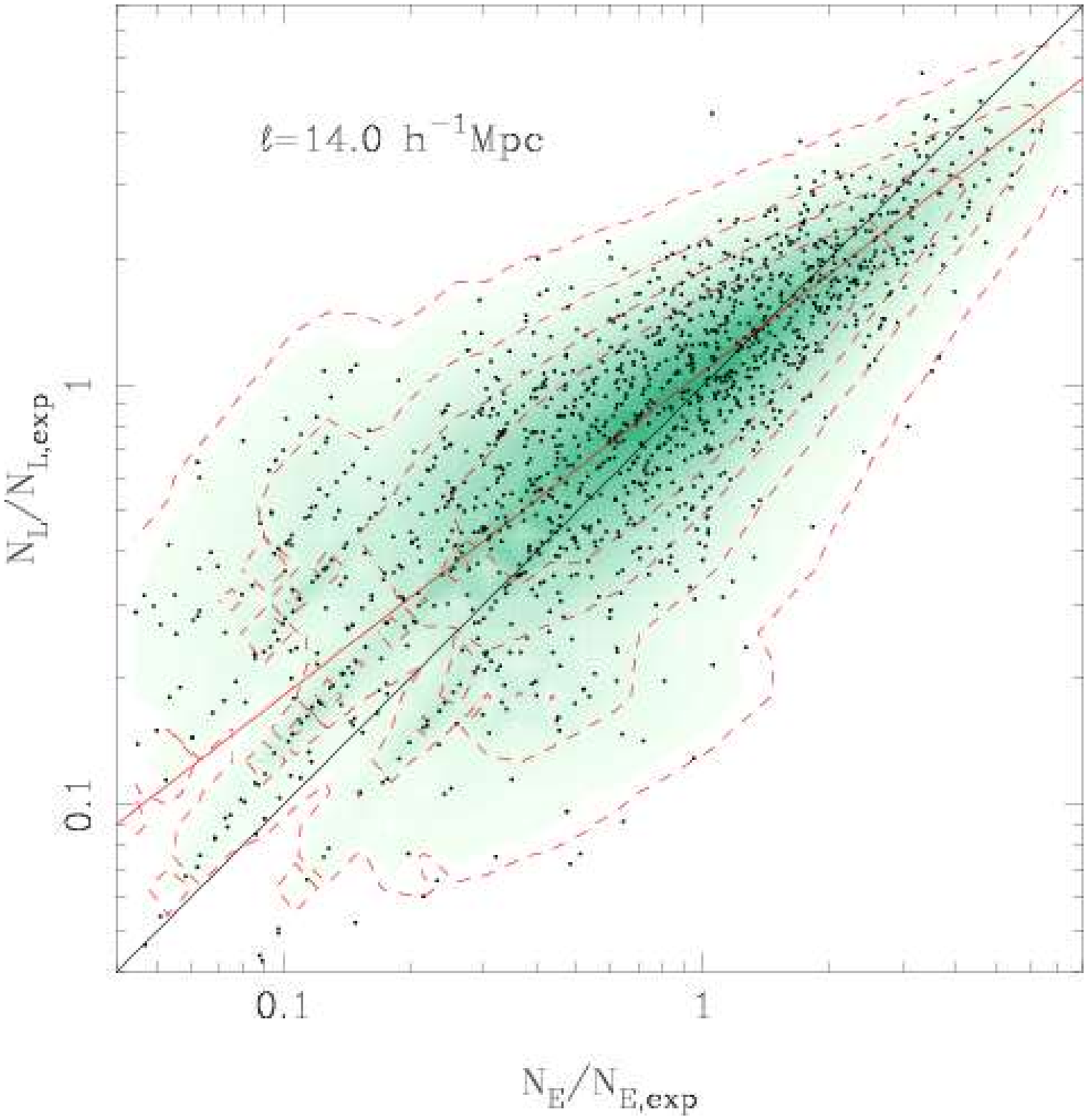}
\end{minipage}\hfill
\begin{minipage}[t]{0.33\textwidth}
\includegraphics[angle=-90, width=0.97\textwidth]{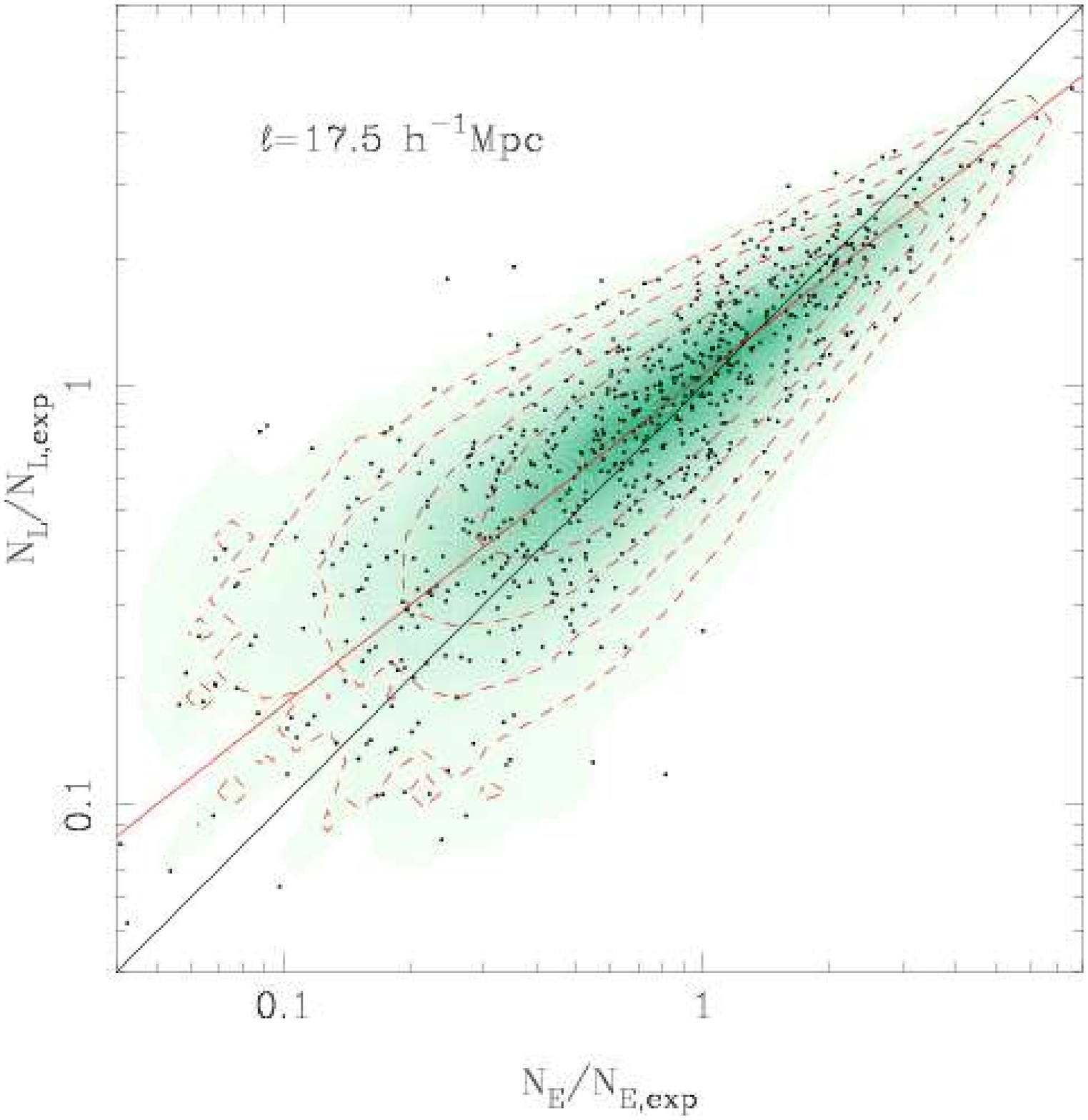}
\end{minipage}
\begin{minipage}[t]{0.33\textwidth}
\includegraphics[angle=-90, width=0.97\textwidth]{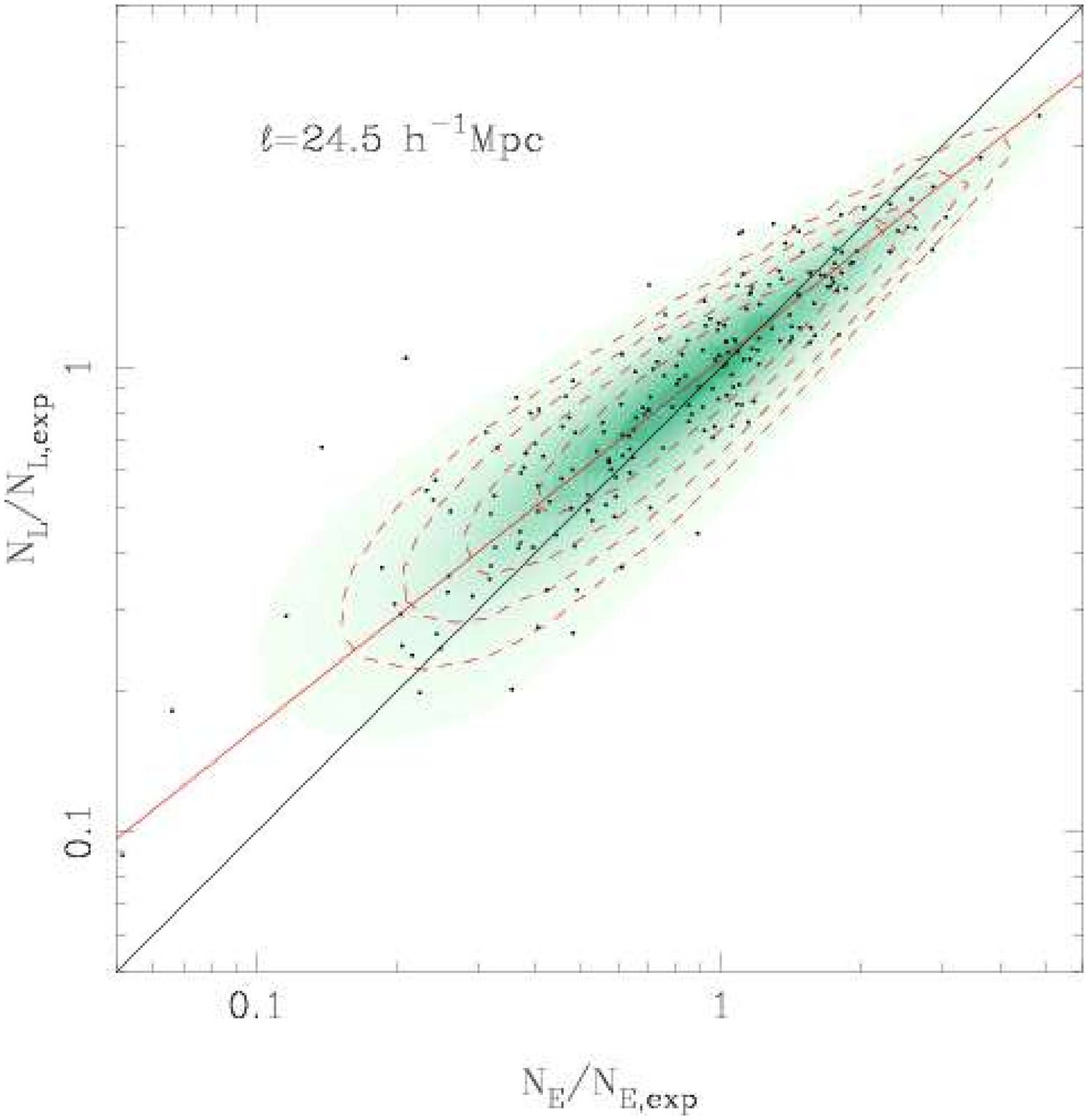}
\end{minipage}
\begin{minipage}[t]{0.33\textwidth}
\includegraphics[angle=-90, width=0.97\textwidth]{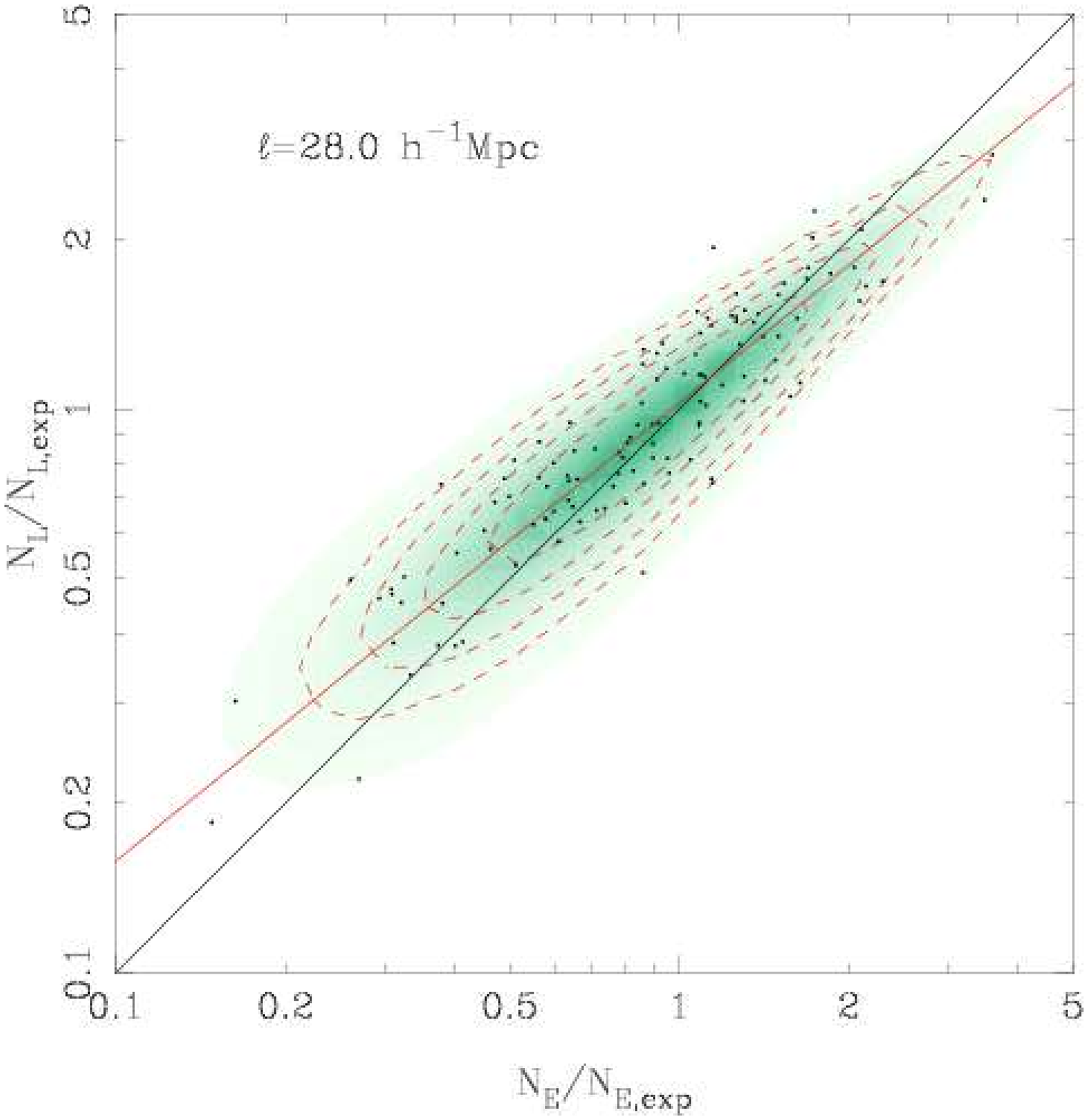}
\end{minipage}\hfill
\begin{minipage}[t]{0.33\textwidth}
\includegraphics[angle=-90, width=0.97\textwidth]{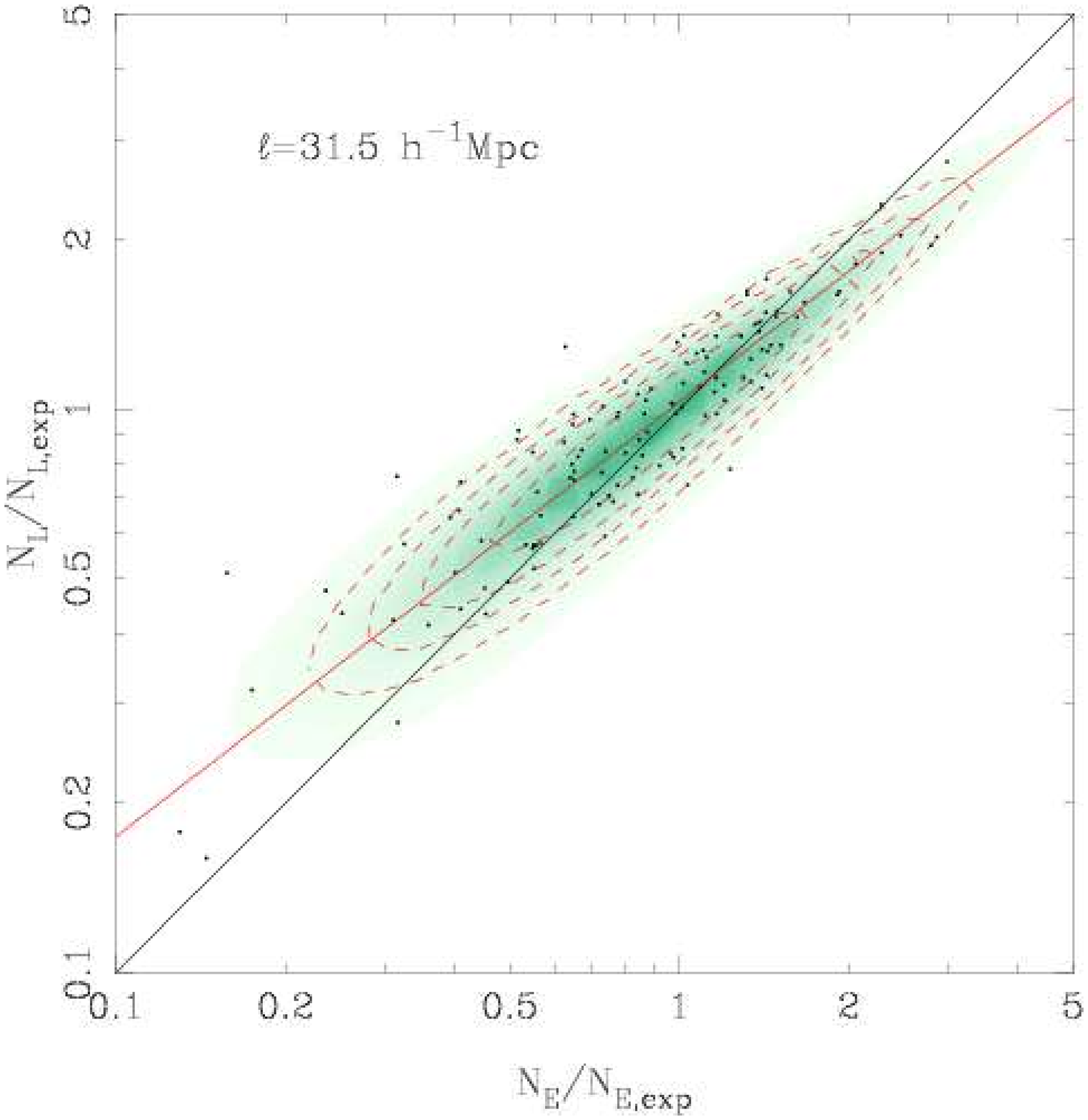}
\end{minipage}
\begin{minipage}[t]{0.33\textwidth}
\includegraphics[angle=-90, width=0.97\textwidth]{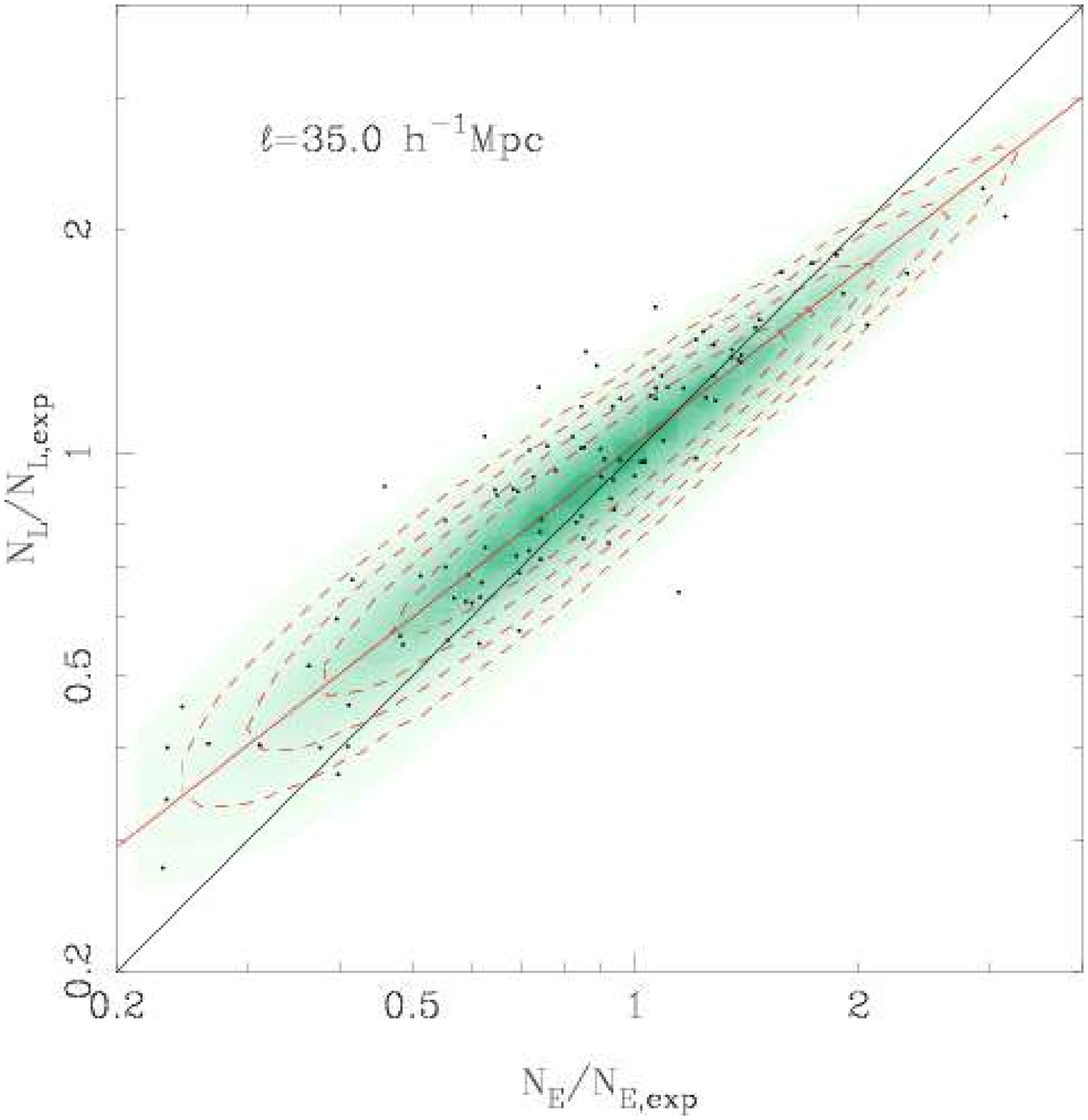}
\end{minipage}
\caption{Contour plots of the joint counts in cells compared to the best fit power-law bias models for the SGP region over a range of cell sizes, from left--right and top--bottom: $\ell=14,17.5,24.5,28,31.5,35 h^{-1}$Mpc. The colour scale and contour levels are derived from Monte Carlo realizations as previously and the dashed contours are at 50\%,70\%,85\% and 93\% significance levels. The black solid line indicates a mean relative bias of 1 and the red line shows the mean relative bias for the model.}
\label{more_conts}
\end{figure*}

Examples of the joint counts in cells for $\ell=21h^{-1}$Mpc compared
to the best fit linear and power-law bias models are shown in
Fig.~\ref{cont_plot}, and illustrations of the power-law bias fits at a
range of scales are shown in Fig.~\ref{more_conts}. The points show
the actual counts in cells. The colour scale and contour levels show the expected
distribution of cell counts, which we have generated using a large number of Monte-Carlo realizations
of the bias model using the same expected counts as the data. Poisson effects are
responsible for the uneven contours on smaller scales; these effects
can also be seen in the data.

\begin{figure}[h]
\begin{center}
\includegraphics[angle=-90, width=0.45\textwidth]{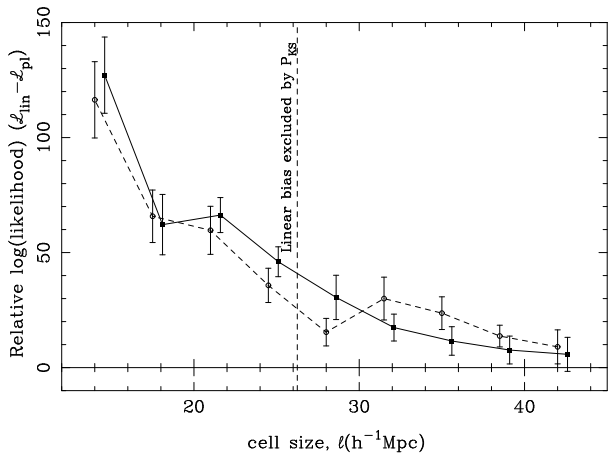}
\end{center}
\caption{The relative likelihoods of the two deterministic bias models when fit to the joint counts-in-cells distribution. A value of 0 implies that both models are equally good fits to the data, positive values indicate that the power-law model is a better fit. Results are shown for the NGP (black squares) and SGP (open circles). The dashed line shows the limit in $\ell$ beyond which a linear bias models is not ruled out by the KS tests. The results, and the error bars shown, are obtained from fits to cell divisions shifted from the original cell division in same manner as was used for the Tegmark tests in Section~\protect\ref{Teg99}.
}
\label{compare}
\end{figure}

To test if the models are acceptable fits to the data we have applied a KS test to the 1-d distributions of the late-type galaxies for these bias models, i.e.~to the projection on the late-type axis of the two dimensional distributions shown in Figs~\ref{cont_plot} \& \ref{more_conts}. The values of $P_{\rm KS}$ for the linear and power-law bias models are shown in the final two columns of Table~\ref{bias_table}; linear bias is excluded for $\ell<28h^{-1}$Mpc cells whereas a power-law bias model is not ruled out for any of the scales considered. Fig.~\ref{compare} shows the relative likelihood $\mathcal{L}_{\rm lin} - \mathcal{L}_{\rm pl}$ of the two models; the power-law bias model is clearly a better fit on smaller scales, although the difference between the goodness of fit of the models decreases with scale, as one would expect from the theoretical prejudice that linear bias should be a good approximation on large scales.

\subsubsection{Non-linear \& stochastic bias}

\begin{table*}
 \centering
 \begin{minipage}{150mm}
  \caption{Average biasing parameters for the best-fitting power-law bias models assuming $b_{\rm var}$ from the individual lognormal model fits to early and late types. $b_{1,{\rm lin}}$ and $b_{1,{\rm PL}}$ are the maximum likelihood results of Section \protect\ref{bias} using realistic error bars and \^b, \~b and $\sigma_{b}$ are calculated from Eqs.~\protect\ref{b_moments}~\&~\protect\ref{sigma_b}. The KS test probabilities for the linear and power-law bias models are also shown.
}
\begin{center}
\begin{tabular}{ccccccccc}
\hline
$\ell$ ($h^{-1}$Mpc) & $b_{1,{\rm lin}}$ & $b_{1,{\rm PL}}$ & \^b & \~b & $b_{\rm var}$ & $\sigma_{b}$ & $P_{\rm KS}$(linear bias) & $P_{\rm KS}$(PL bias) \\	
\hline
% 14  & 0.753$\pm$0.009 & 0.68$\pm$0.04 & 0.69$\pm$0.05 & 0.73$\pm$0.02 & 0.24$\pm$0.2  & 1.3e-5  & 0.269 \\
% 21  & 0.74$\pm$0.01 & 0.69$\pm$0.06 & 0.70$\pm$0.07 & 0.73$\pm$0.04 & 0.21$\pm$0.3    &   0.004   & 0.412 \\
% 28  & 0.74 $\pm$0.01 & 0.72$\pm$0.09 & 0.7$\pm$0.1 & 0.74$\pm$0.06 & 0.13$\pm$0.7     &    0.688   & 0.712 \\
% 35  & 0.73$\pm$0.01 & 0.7$\pm$0.1 & 0.7$\pm$0.1 & 0.74$\pm$0.07 & 0.20$\pm$0.5        &    0.320   & 0.760 \\
% 42  & 0.71$\pm$0.02 & 0.7$\pm$0.1 & 0.7$\pm$0.1 & 0.75$\pm$0.09 & 0.23$\pm$0.5        &    0.723   & 0.957 \\
14 & 0.83$\pm$0.02 & 0.76$\pm$0.02 & 0.68$\pm$0.06 & 0.7$\pm$0.1 & 0.73$\pm$0.02 & 0.2$\pm$0.3  & 1.3e-5  & 0.269   \\
21 & 0.82$\pm$0.03 & 0.75$\pm$0.03 & 0.69$\pm$0.09 & 0.7$\pm$0.1 & 0.73$\pm$0.04 & 0.2$\pm$0.5  &   0.004   & 0.412  \\
28 & 0.79$\pm$0.08 & 0.75$\pm$0.08 & 0.7$\pm$0.2 & 0.7$\pm$0.3 & 0.74$\pm$0.06 & 0.1$\pm$2   &    0.688   & 0.712 \\
35 & 0.8$\pm$0.1 & 0.7$\pm$0.1 & 0.7$\pm$0.2 & 0.7$\pm$0.4 & 0.75$\pm$0.07 & 0.2$\pm$2      &    0.320   & 0.760 \\
42 & 0.7$\pm$0.2 & 0.7$\pm$0.2 & 0.7$\pm$0.5 & 0.7$\pm$0.9 & 0.7$\pm$0.1 & 0.2$\pm$3       &    0.723   & 0.957 \\   
\hline
\label{bias_table}
\end{tabular}
\end{center}
\end{minipage}
\end{table*}

In the case of linear and deterministic bias, all three bias
parameters described at the start of the section (\^b\ ,\~b and
$b_{\rm var}$), are equal to the parameter $b_{1}$ in our model
(Eq.~\ref{lin_bias}). We tabulate the values of \^b and \~b for our
best-fitting power-law bias model below. In the absence of
stochasticity we would have $b_{\rm var}=\tilde{b}$, but since we
already have an estimated value for $b_{\rm var}$ from the independent
fitting of lognormal models to the early- and late-type counts in
cells, we can instead ask what value the stochastic bias parameter,
$\sigma_{b}$ (Eq.~\ref{sigma_b}), should take under the rather strong
assumption that a power-law bias completely describes any
nonlinearity. The average biasing parameters under this assumption are
summarized in Table~\ref{bias_table}. $\sigma_{b}$ is generally
$\sim0.2$, although the large errors on $b_{\rm var}$ mean that we
cannot claim to require excess stochasticity above Poisson noise
within our errors. A detailed model of stochastic relative bias is
discussed by Wild \etal\ (in preparation); our results are consistent with the
more accurate measurements presented in that paper. 
Similarly, the non-linearity quantified by \~b/\^b from our
measurements is entirely consistent with that measured by Wild et al. 
Szapudi \& Pan
(2003) describe an interesting technique, based on the cumulative
distribution functions, which can in principle recover the full
non-linear bias function. This would enable a model independent
measurement of nonlinearity and stochasticity, although in unmodified
form the technique is not applicable to a flux-limited sample.

\section{Discussion}\label{concl}
In this paper we have presented a number of measurements of the relative bias between early- and late-type galaxies in the 2dFGRS derived using the counts in cells and the joint counts in cells for the separate galaxy populations. The behaviour of individual estimators for the linear relative bias parameter as a function of scale, as well as the relationship between different estimators of the linear relative bias parameter as a function of scale both have important implications for the scale dependence, nonlinearity and stochasticity of the relative bias between early- and late-type density fields. We have also used a power-law bias model as the simplest model including non-linear effects and demonstrated the characteristics of the best fit power-law bias model as a function of scale.
\subsection{Variances and the one-point distribution function}
 We have presented the variance of the counts in cells using the
 maximum-likelihood technique of Efstathiou et al.~(1990), which we
 have shown is subject to a significant bias when dividing the data
 into  redshift shells of low volume. We have shown that the method can be corrected for this integral constraint bias using the approximation of Hui \& Gazta\~naga (1999).

The one-point distribution of the counts in cells for early- and
late-type galaxies, and the distribution for all $\eta$-typed
galaxies, has been fit by lognormal models, using a maximum-likelihood
technique. The variances found using this technique are significantly
biased on small scales when empty cells are included in the analysis,
and we have been able to measure reliable variances only by fitting to
counts in cells with empty cells removed. We have corrected our
results on small scales to compensate for the inevitable bias
resulting from the removal of empty cells. We find that the lognormal
model is in general an adequate fit to the distribution functions, as
measured by a Kolmogorov--Smirnov test. However the values for the
variance implied by the best fit model parameters are slightly high in
comparison with both predictions from the correlation functions and
relative to the direct counts-in-cells variance measurements presented
in this paper. The fact that this bias can be corrected by introducing
a weighting scheme giving more weight to regions of higher density
contrast suggests that the lognormal model is a relatively crude
approximation to the true distribution. It is likely that a
generalized lognormal model, such as the `skewed' lognormal model
(SLNDFk) (Colombi 1994; Ueda \& Yokoyama 1996), would be a better
approximation. Unfortunately, the SLNDFk cannot be used in our maximum
likelihood approach since it is not positive definite, and therefore is
not strictly speaking a distribution function.

\subsection{Comparison of relative bias parameters}
\label{last_bit}

\begin{table}
  \caption{Average bias parameters over all scales from $\ell>14h^{-1}$Mpc for all of the measurements presented in the paper. Error bars are derived assuming measurements for adjacent bins in $\ell$ scales are correlated.}

\begin{center}
\begin{tabular}{ccc}
\hline
Bias measurement & NGP & SGP \\	
\hline
$1/b_{\rm var}$ (from Efstathiou $\sigma(\ell)$) & 1.24$\pm$0.06 & 1.26$\pm$0.04 \\
$1/b_{\rm var}$ (from $\sigma_{\rm LN}$ fits) & 1.28$\pm$0.05 & 1.27$\pm$0.04 \\
$1/b_{1,{\rm lin}}$ (maximum likelihood) & 1.27$\pm$0.04 & 1.17$\pm$0.04 \\
$f\approx1/b_{1,{\rm lin}}$ (Tegmark test) & 1.28$\pm$0.03 & 1.16$\pm$0.03 \\
$1/b_{1,{\rm PL}}$ & 1.36$\pm$0.05 & 1.29$\pm$0.04 \\
\hline
\label{av_bias}
\end{tabular}
\end{center}
\end{table}

We present in Table~\ref{av_bias} a comparison of the relative bias parameters from all of the measurements presented in the paper. We have averaged the bias measurements for all scales with $\ell>14h^{-1}$Mpc which we expect to be unaffected by biases from empty cells. The error bars on each average are obtained assuming that the measurements in adjacent bins in $\ell$ are perfectly correlated, which is a better approximation than assuming the measurements on separate scales are independent. Where relevant we have also used the more realistic error bars obtained from our Rayleigh-L\'evy flight models.

As previously noted, the results for $1/b_{\rm var}$ are consistent between regions and also consistent between measurements from direct variance estimation and fitting lognormal models to the one-point distribution.

Comparing the two estimates of the linear relative bias parameter $1/b_{1,{\rm lin}}$, from the maximum likelihood method and from the Tegmark test, we find in both cases a significant discrepancy between NGP and SGP regions. The magnitude of this discrepancy is around 2-$\sigma$. On the other hand the power-law bias measurements are approximately consistent between regions at a value of $b_{1,{\rm PL}}$ which is further from unity. As we noted in Section~\ref{bias} the assumption of linear bias when fitting to joint counts in cells which contain a significant degree of nonlinearity pushes the best fit relative bias closer to unity. This effect was also noted by Wild \etal\ (in preparation). It is likely that the apparent discrepancy between NGP and SGP linear bias parameters is also partly an artefact produced when non-linear joint distributions are fit with a linear bias model.

\subsection{Scale dependence of the relative bias}
In general, the relative bias is expected to be scale dependent on small scales ($r\lesssim r_{0}$). The scale at which the bias relation becomes scale independent depends on the scales over which the biasing mechanism(s) operates. Non-local bias models (Bower et al.~1993, Matsubara 1999) are those on which the physical processes acting to produce the bias act on scales larger than those defined by the movement of massive particles, for example those models where radiation from QSOs has a significant effect. Local bias models (e.g.~Narayanan, Berlind \& Weinberg, 2000) are those which are defined by some property of the local matter field, for example its density.

Narayanan, Berlind \& Weinberg (2000) determine the variation with scale of a number of local and non-local bias models applied to N-body simulations. A general conclusion of this work is that local bias models are generically unable to influence the biasing relation on scales greater than $r=8h^{-1}$Mpc, which corresponds to $\ell\approx12h^{-1}$Mpc in this work. Although there does appear to be some variation of the best fit linear bias parameter on scales $\ell\gtrsim15h^{-1}$, when we factor in the larger error bars derived from models including selection function variation across cells in a more realistic way, the significance of any variation becomes negligible. Even if the variation in the linear bias parameter were significant it would not necessarily imply scale dependence of the bias since there is no significant scale dependence of the best fit power-law bias parameter. This illustrates the interdependence of non-linear, non-local and stochastic biasing effects. We conclude that any non-local contribution to the relative bias cannot be a dominant effect on large scales.

A special case of local relative bias which was considered in detail by Narayanan, Berlind \& Weinberg (2000) is a local morphology density relation, of the type measured in the local environment of clusters and groups by e.g.~Postman \& Geller (1984). A general conclusion for the relative bias produced by such a local effect is that the constant bias factor to which the scale dependent bias asymptotes on large scales is not equal to unity; assigning galaxy types based on local density produces a difference in clustering strength of the different galaxy types on all scales. Our results are fully consistent with this picture, which leads on to the question of whether the relatively well studied morphology-density relation can be held solely responsible for the relative bias measured in the 2dFGRS.

\section*{Acknowledgments}

The 2dFGRS was made possible through the dedicated efforts of the staff of the Anglo-Australian Observatory, both in creating the 2dF instrument and in supporting it on the telescope. JAP and OL thank PPARC for their Senior Research Fellowships.

\setlength{\bibhang}{2.0em}

\label{lastpage}

\end{document}